\documentclass[twocolumn]{aastex631}

\usepackage{CJK}
\usepackage{soul}

\usepackage[normalem]{ulem}

\newcommand\St{\,{\rm St}}

\definecolor{emerald}{RGB}{0,155,155}

\begin{document}
\begin{CJK*}{UTF8}{gbsn} 

\title{Observability of substructures in planet-forming disk in (sub)cm wavelength with SKA and ngVLA}

\author[0000-0003-3728-8231]{Yinhao Wu (吴寅昊)}
\affiliation{School of Physics and Astronomy, University of Leicester, Leicester LE1 7RH, UK; \href{mailto:email@domain}{yw505@leicester.ac.uk}}

\author[0000-0002-9442-137X]{Shang-Fei Liu (刘尚飞)}
\affiliation{School of Physics and Astronomy, Sun Yat-sen University, Zhuhai 519082, China;
\href{mailto:email@domain}{liushangfei@mail.sysu.edu.cn}} 

\author[0000-0003-2948-5614]{Haochang Jiang (蒋昊昌)}
\affiliation{European Southern Observatory, Karl-Schwarzschild-Str 2, 85748 Garching, Germany;}
\affiliation{Department of Astronomy, Tsinghua University, Haidian DS 100084, Beijing, China}

\author[0000-0002-6166-2206]{Sergei Nayakshin}
\affiliation{School of Physics and Astronomy, University of Leicester, Leicester LE1 7RH, UK; \href{mailto:email@domain}{yw505@leicester.ac.uk}}


\begin{abstract}
Current imaging observations of protoplanetary disks using ALMA primarily focus on the sub-millimeter wavelength, leaving a gap in effective observational approaches for centimeter-sized dust, which is crucial to the issue of planet formation. The forthcoming SKA and ngVLA may rectify this deficiency. In this paper, we employ multi-fluid hydrodynamic numerical simulations and radiative transfer calculations to investigate the potential of SKA1-Mid, ngVLA, and SKA2 for imaging protoplanetary disks at sub-cm/cm wavelengths. We create mock images with ALMA/SKA/ngVLA at multi-wavelengths based on the hydrodynamical simulation output, and test different sensitivity and spatial resolutions. We discover that both SKA and ngVLA will serve as excellent supplements to the existing observational range of ALMA, and their high resolution enables them to image substructures in the disk's inner region ($\sim$ 5 au from the stellar). Our results indicate that SKA and ngVLA can be utilized for more extended monitoring programs in the centimeter waveband. While in the sub-centimeter range, ngVLA possesses the capability to produce high-fidelity images within shorter observation times ($\sim$ 1 hour on source time) than previous research, holding potential for future survey observations. We also discuss for the first time the potential of SKA2 for observing protoplanetary disks at a 0.7 cm wavelength.
\end{abstract}


\section{Introduction}\label{sec:1}
Planet formation and its intricate relation with protoplanetary disks have remained at the forefront of astrophysical research for decades. One of the most pivotal pieces of this puzzle lies in the substructures of these disks, especially in the form of gaps and rings. The Atacama Large Millimeter/Submillimeter Array (ALMA) has been invaluable in observing these substructures in many systems \citep[e.g.,][]{ALMA2015,HD169142-2017,Clarke-2018,Dipierro-2018,DSHARP-I,Long-2019,Long-2020,Long-LkCa15-2022,Long-2023}, thanks to its high-resolution capabilities.

These substructures can provide crucial information about the dynamic interplay between forming planets and their natal environments. While several mechanisms, such as changes in dust properties around ice lines \citep{Zhang-2015,Okuzumi-2016} or inherent (magneto-)hydrodynamic (MHD) instabilities in disk \citep{Flock-2015,Suriano-2018,Riols-2019,Riols-2020,MHD-wind-Elbakyan,Wu-Chen-Jiang_2023}, have been postulated to account for these patterns, a prevailing hypothesis is that they are a direct consequence of interactions between the disk and embedded, still-forming planets \citep[see][for reviews]{Kley-Nelson-2012-araa,Baruteau-PPVI,Andrews-araa-2020}.

As a young massive planet orbits its host star, it exerts gravitational forces on the surrounding disk material, potentially carving out a gap in its path. Notably, localized continuum excesses \citep{Keppler-2018,Tsukagoshi-2019,Nayakshin-2020} and discernible kinematic deviations in the disc's molecular gas—attributed to the influence of an embedded planet \citep{Teague-2018,Pinte-2019,Pinte-2020,Pinte-2023} —have further underscored this interpretation in recent years.

While there's an abundance of observational evidence pointing to planet-induced substructures in discs, the leap from millimeter(mm)-size dust particles to kilometer-size planetesimals still a challenge. Collisions at the mm-scale often result in fragmentation or a rebound effect, rather than the anticipated merging \citep[see][for review]{Blum-2008}. Additionally, the solid material within the disc is subject to aerodynamic drag from surrounding gases, which saps angular momentum, propelling these materials swiftly towards the central star \citep{Weidenschilling1977}. Although some theories suggest accelerated growth mechanisms—like the streaming instability \citep[e.g.,][]{Youdin-2005,Johansen-2007,Wu_Lin-2024}, its evident that the transition from mm-sized dust to centimeter(cm)-sized pebbles requires further elucidation to unlock the intricacies of planet formation. Notably, at mm wavelengths, it's primarily mm-sized dust grains that dominate dust opacity \citep{Draine-1984,Semenov-2003,Draine-2006}. Hence, ALMA, spanning a wavelength range from 0.3 to 3 mm, is adept at capturing emissions from roughly mm-sized grains. Yet, its sensitivity falters when observing emissions from larger, cm-sized dust grains.

In any sense, observations at longer wavelengths are crucial to improve the chance of revolving the high optical depth region, and the only way to robustly decide the size distribution of pebbles in the corresponding wavelength range. Yet, currently, the Janksy Very Large Array (VLA) is the only facility capable of providing reasonable sensitivities and angular resolution at cm wavelength, whose observation could be only archived in a handful of brightest disks. 

Upcoming facilities, like the Square Kilometer Array's mid-frequency component (SKA1-Mid) \citep{Braun-2015-SKA,Braun-2019-SKA} and its subsequent phase (SKA2), along with the next-generation Very Large Array (ngVLA) \citep{Murphy-2018-ngVLA}, offer significantly improved sensitivity and resolution expected at (sub)cm bands.Only with these, larger-size grain populations and their distribution in protoplanetary disks can be resolved in the planet-forming au scales \citep{Ricci_2018,Ilee_2020,Harter2020,Ricci-2021,Ueda-2022}.

In this work, we simulate the potential of SKA1-Mid, SKA2, and ngVLA in observing the substructures, particularly gaps and rings, created by gap-opening planets in protoplanetary disks. By juxtaposing their capabilities with the existing benchmark set by ALMA Band 1, we endeavor to provide a clearer picture of how future observations can shed light on the age-old questions surrounding planet formation, particularly through the lens of dust grain distribution in protoplanetary disks.

Our paper is organized as follows: In \S \ref{sec:2}, 
we introduce the numerical setup for our hydrodynamic simulations, radiative transfer calculations, and parameter choices for observational mock imaging. 
In \S \ref{sec:3}, we described the mock images for ALMA band 1, SKA, and ngVLA at (sub))cm wavelength. We present a summary of our findings and engage in a comprehensive discussion regarding the implications of our results in \S \ref{sec:4}.

\section{Numerical Setup}\label{sec:2}
In this section, we introduce the methods that we used to simulate the high-resolution observation in radio wavelengths.

\subsection{Hydrodynamic Simulations}\label{sec:2.1}
In this study, we used grid-based code \texttt{FARGO3D} \citep{FARGO3D} to do 2D \{$r,\phi$\} hydrodynamical planet-disk simulations. This code is the successor of the public \texttt{FARGO} code \citep{FARGO}, in the latest version, it added a multi-fluid module \citep{FARGO3D-multifluid}, we can consider the dynamics of gas and dust in a disk with one or more embedded planet. This code replaced dust-particle size with a constant Stokes Number $\St$ to describe interacting with the gas. However, in this study, we used fixed dust sizes to describe the drag force between fluids. 

We adopt the same solar-like system numerical simulation settings as the previous works \citep{Ricci_2018}. In this study, we adopted the following initial radial profile of gas surface density:
\begin{equation}\label{eq:sigma_g}
\Sigma_{gas}(r, t=0) = \Sigma_{0}\left(\frac{r}{r_c}\right)^{-\gamma} exp\left[-\left(\frac{r}{r_c}\right)^{2-\gamma}\right],
\end{equation}
this equation is the self-similar solution for the evolution of a viscous accretion disk, when the viscosity $\nu(r) \propto r^{\gamma}$ \citep{Lynden-Bell_1974}. For equation \ref{eq:sigma_g}, $r$ is the radius in the disk, $r_c$ is a truncation radius, and normalization constant $\Sigma_{0}=1\times10^{-4} M_{\star}/r_0^2$ is a parameter used to describe the initial total gas mass in the disk. In our simulations, we set values of $\gamma = 0.8$, $r_c = 17$ au and the total disk mass $M_{disk}=8\times10^{-3}M_{\odot}$.

For the initial dust surface density, we set
\begin{equation}\label{eq:sigma_d}
\Sigma_{dust}(r, t=0) = \epsilon \times \Sigma_{gas}(r, t=0),
\end{equation}
where dust-to-gas mass ratio $\epsilon = 0.01$. We distributed 11 different sizes of dust ranging from 7 $\mu$m to 1 cm in our simulations. For each size of dust particles, their mass follows grain-size ($a$) distribution $n(a) \propto a^{-q}$ \citep[e.g.,][]{Ricci_2010}, where $q = 3.5$. We set the internal density of dust to be $1.2~g/cm^{3}$ for all sizes. 

The mass of the central star is set $M_\star = 1~M_{\odot}$. We set the geometrical aspect ratio of the disk, $H/R$, to 
\begin{equation}\label{eq:h}
    h(r) = h_0 \times \left(r/r_0\right)^{f}, 
\end{equation}
where $h_0 = H/r = 0.05$, $f = 0.25$ is a constant appropriate for the disc midplane temperature profile $T(r) \propto r^{-1/2}$, $r_0$ is the planet orbital radius. For the gas viscosity, we utilize the $\alpha$ prescription of \cite{SS_1973} and adopt two values of the $\alpha$ parameter, $10^{-3}$ and $10^{-5}$. 

In this study, we focus on the disk structure of massive planets, and we only consider the same orbital radius,  We let the orbital radii be kept fixed during each simulation. In our simulations, we put a Jupiter mass planet (planet-to-star mass ratio is $q = 1\times 10^{-3}$) for gap-opening. We run the simulations for 1500 orbital periods, and the planets’ orbital radii are set to 5 au in our simulations, for our orbital radii, this is corresponding to 16700 yr. 

Our numerical grids extend from 0.1 $r_0$ (0.5 au) to 12 $r_0$ (60 au) in the radial direction and 0 to $2\pi$ in the $\theta$ direction. We utilized 1536 grids in the radial direction and 1024 grids in the $\theta$ direction.


\subsection{Radiative Transfer Calculations}\label{sec:2.2}
We utilized the publicly available \texttt{RADMC-3D} code \citep{RADMC-3D-2012} to conduct our dust radiative transfer calculations. Then to process our \texttt{FARGO3D} simulations for use as inputs in \texttt{RADMC-3D}, and to compute the final beam-convolved synthetic maps of dust continuum emission, we employed the publicly available Python code \texttt{fargo2radmc3d} \citep{fargo2radmc3d-dust}. 

The configuration of the dust radiative calculations closely follows the methodology described in \citep{Baruteau21}, this method has been applied in multiple recent studies \citep[e.g.,][]{Wu2023,Wu-Chen-Jiang_2023}. Notably, the dust temperatures employed in \texttt{RADMC-3D} align with the temperatures utilized in the hydrodynamical simulations. Here, we only briefly list the main parameters used in our study.

To extend the 2D surface density in hydrodynamical simulation to 3D spherical grids for radiation transferring calculation, we consider vertical hydrostatic equilibrium and set the dust scale heights to
\begin{equation}
    h_{\rm i,dust}=h(r)\times\left(\frac{\alpha}{\alpha+\St_{\rm i}}\right)^{1/2},
    \label{dust-scale-height}
\end{equation}
here $i=1...11$ and $\St_{i}$ is the average Stokes number for the $i^{\rm th}$ dust fluid. The spherical grid encompasses approximately $\pm$3H over the midplane, consisting of 40 grids with a non-uniform spacing. Henyey-Greenstein anisotropic dust scattering \citep{Savage-Mathis-1979} is incorporated into the calculations. The dust size distribution is described in $\S$\ref{sec:2.1}.

The opacities for dust absorption and scattering are computed using Mie theory, and are produced by \texttt{OpTool} \citep{optool-2021}, the grain composition is the same as DSHARP composition \citep{DSHARP-V,DSHARP-VII}, which contains the total opacities of absorption and scattering for a given wavelength and grain size. The specific intensity of the dust continuum emission is computed with $10^8$ photon packages used for ray tracing.

\begin{figure*}
\centering
\includegraphics[width=1\hsize]{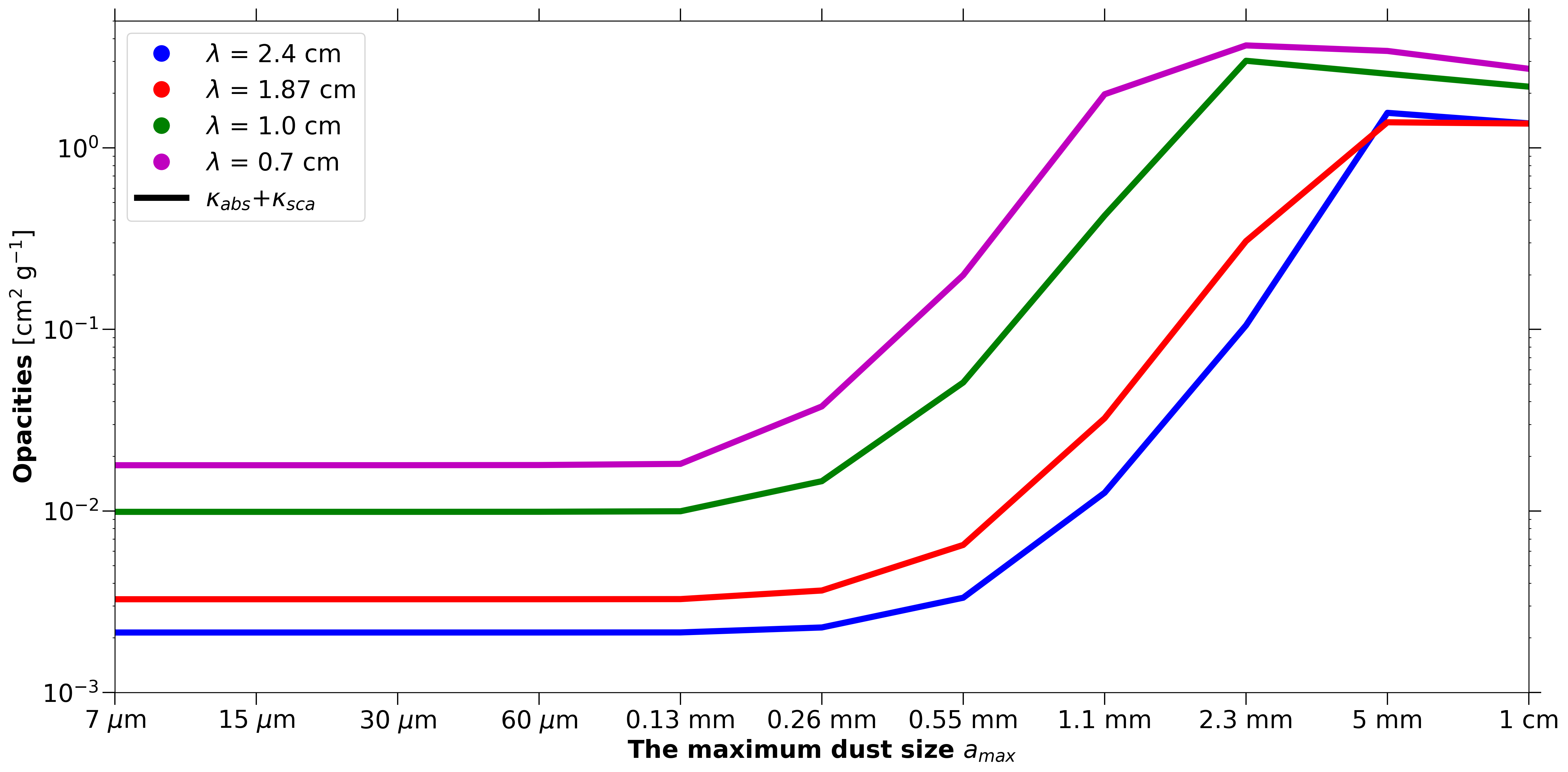}
\caption{The total (absorption + scattering) optical depth ($\kappa_{abs}+\kappa_{sca}$) versus the maximum dust, are used in our radiation transfer calculations). Different colors indicate different observation wavelengths.}   
\label{fig:opac}
\end{figure*}

In Fig.\ref{fig:opac}, we plot the relationship between the total optical depth ( absorption $+$ scattering) and the maximum size dust $a_{max}$ at different observation frequencies (wavelengths). This shows a huge potential of the observations in longer $\lambda$ to constrain the dust size distribution.

We set the disk distance is 150 parsecs far from antennas, which distance encompasses hundreds of disks in regions such as Taurus \citep{Taurus-2007}, Ophiuchus \citep{Ophiuchus-2017}, Chamaeleon \citep{Chamaeleon-1997}, and Lupus \citep{Lupus-2008}. We assume the stellar effective temperature of 6000~K with star radius is 1.5 $R_\odot$. Taking radiative transfer at the 0.7 cm wavelength as an example, our total disk flux is 0.26 mJy (for disk model with $\alpha=10^{-3}$) and 0.2 mJy (for disk model with $\alpha=10^{-5}$).

\subsection{Simobserve}\label{sec:2.3}
For each hydrodynamic simulation, we make synthetic maps to mock ALMA, SKA or ngVLA images at varied wavelengths. We utilized the \texttt{SIMOBSERVE} task within the \texttt{Common Astronomy Software Application} (\texttt{CASA 6.4.9}) package to transform synthetic images into visibility data sets in the Fourier domain. 

For configuring the ALMA and ngVLA observations, we employed the alma.out28.cfg and ngvla-main-revC.cfg files provided within the \texttt{CASA} package. The ALMA configuration includes baselines extending up to 16 km, while the ngVLA's reference design features 214 antennas, each 18 meters in diameter, with baselines reaching approximately 1000 km. For the ngVLA, we utilized \texttt{SETNOISE} for noise processing. And for \texttt{TCLEAN} task, we set psfcutoff to 0.75 because our PSFs with a large skirt. 

For SKA1-Mid, we utilized the beam size and weighted continuum sensitivity calculated by the \texttt{SKAO Sensitivity Calculator}\footnote{\href{https://sensitivity-calculator.skao.int/mid}{https://sensitivity-calculator.skao.int/mid}}, and then created synthetic maps using \texttt{fargo2radmc3d}. For observations at 12.5 GHz on one hour source time, the beam size of SKA1-Mid is $0.025 \times 0.022$ arcsec, with a sensitivity of 14.5 μJy/beam. Since SKA2 remains conceptual at this stage, we adopted two design values for the beam size ($0.01 \times 0.01$ arcsec and $0.019 \times 0.019$ arcsec) provided by \cite{Braun-2019-SKA} and a reference Weighted continuum sensitivity of 3.2 μJy/beam for one hour of observation time to create synthetic maps.

For this study, we conducted our simobserve at two different frequencies for SKA. One frequency of 12.5 GHz (wavelength $\sim2.4$ cm) is positioned at the midpoint of the suggested SKA1-Mid Band 5b frequency range ($8.3-15.3$ GHz). This choice is similar to \cite{Ilee_2020}. The other one is 43.3 GHz (wavelength $\sim0.7$ cm), which is a possible upgrade path of SKA2 \citep[as part of the SKA Observatory development plan,][]{Braun-2019-SKA}. For ngVLA, we took into account three distinct frequency bands, Band 3 (16 GHz, wavelength $\sim1.87$ cm), Band 4 (30 GHz, wavelength $\sim1.0$ cm), Band 5 (43.3 GHz, wavelength $\sim0.7$ cm). As a control, we also made the synthetic maps of ALMA Band 1 (43.3 GHz, wavelength $\sim0.7$ cm). Specifically, the antennas of ALMA Cycle 11 (C43-10). We summarise the array layout definitions in Table \ref{table:2} for reader's convenience. 

\begin{deluxetable*}{lccccc}
  \tablecaption{The label, array names, frequencies, wavelength, angular resolutions, sensitivities in the computation of the synthetic observations. These sensitivity values are specifically applicable to observations conducted with 1 hour of on-source time.
\label{table:2}}
    \tablehead{\colhead{Label} & \colhead{Array Name} & \colhead{Frequency (GHz)} & \colhead{Wavelength (cm)} & \colhead{Angular Resolution (arcsec)} & \colhead{Sensitivity ($\mu Jy/beam$)}
    } 
\startdata
\hline\hline
Fig. \ref{fig:image-SKA1 Mid High} & SKA1-Mid (a) & 12.5 & 2.4 & $0.025\times0.022$ & 14.5 \\
Fig. \ref{fig:image-ngVLA 1.87} & ngVLA Band 3 & 16 & 1.87 & $0.0051\times0.0034$ & 0.23 \\
Fig. \ref{fig:image-ngVLA 1.0} & ngVLA Band 4 & 30 & 1.0 & $0.0031\times0.0025$ & 0.24 \\
\hline
Fig. \ref{fig:image-0.7} & SKA2 (plan, a) & 43.3 & 0.7 & $0.01\times0.010$ & 3.2 \\
Fig. \ref{fig:image-0.7} & SKA2 (plan, b) & 43.3 & 0.7 & $0.019\times0.019$ & 3.2 \\
Fig. \ref{fig:image-0.7} & ngVLA Band 5 & 43.3 & 0.7 & $0.0027\times0.0023$ & 0.32 \\
Fig. \ref{fig:image-0.7} & ALMA Band 1 C43-10 (Cycle 11) & 43.3 & 0.7 & $0.1\times0.1$ & 8.5 \\
\enddata
\end{deluxetable*}


\section{Results}\label{sec:3}
In this section, we display the results of our multi-fluid hydrodynamic simulations for a disk with an embedded planet and the interferometric observations' potential with SKA, ngVLA and ALMA. 

\begin{figure}
\centering
\includegraphics[width=0.49\hsize]{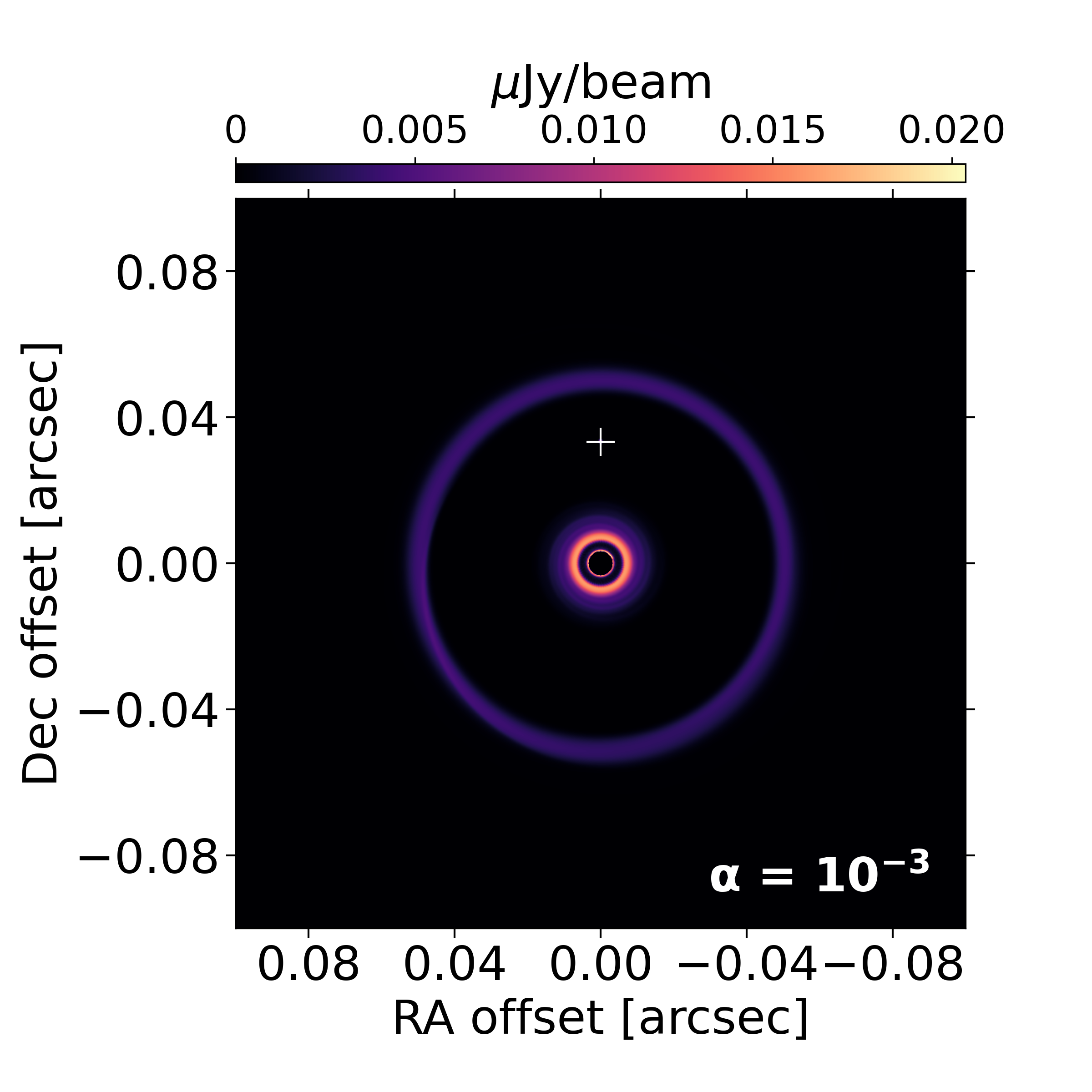}
\includegraphics[width=0.49\hsize]{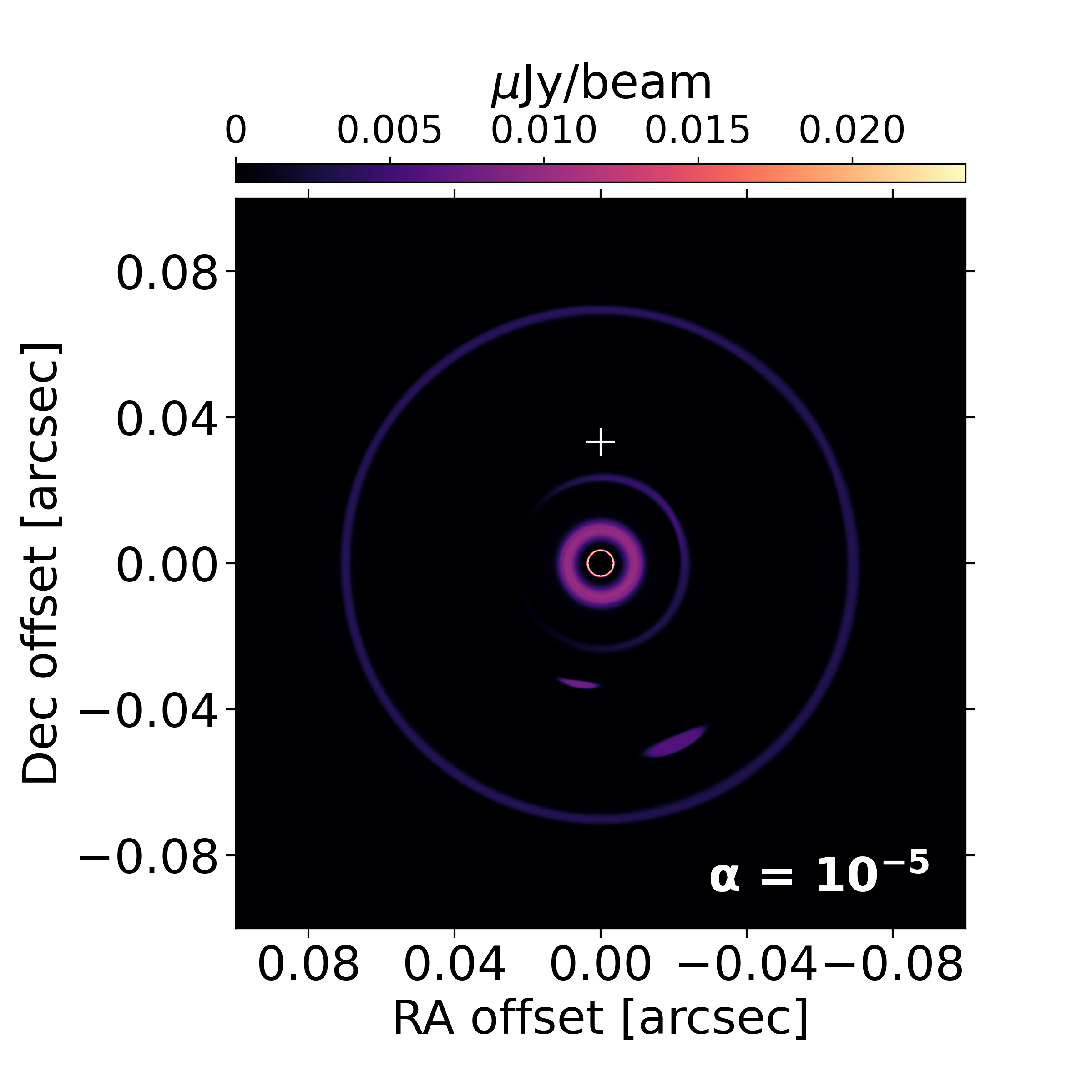}
\caption{The skymodels in units of $\mu$Jy/beam that we used for this study. In each panel, the value of $\alpha$ viscosity of each model is labeled in the lower right corner of each panel. The white ``+" symbol indicates the position of the planet.}\label{fig:hydro}
\end{figure}

In Fig.\ref{fig:hydro}, we present the skymodel for different models that we used for next simobserve. We noticed that model with $\alpha = 10^{-3}$ exhibits a very symmetrical and bright ring with a deep gap. The model with $\alpha = 10^{-5}$, on the other hand, generates a bright vortex in the horseshoe region and asymmetric crescent-like structures in the inner region, as well as wider gaps and narrower rings in the outer region. This is reasonable because its extremely low viscosity will make it easier to open gaps \citep[e.g.,][]{Kanagawa-2016,MHD-wind-Elbakyan}, make secondary inner gap \citep{Dong2017,2018HuangDSHARP2}. And due to Rossby Wave Instability \citep[RWI;][]{RWI-1999,rwi-2001}, it will generate vortex accordingly.

We regard the successful observation of bright rings, dark gaps, and all asymmetric structures presented in the fluid model as critical factors in assessing the antenna's performance and potential.

\subsection{Observation at 2.4 cm Band with SKA1-Mid}\label{sec:SKA1 Mid}
\begin{figure}
\centering
\includegraphics[width=0.49\hsize]{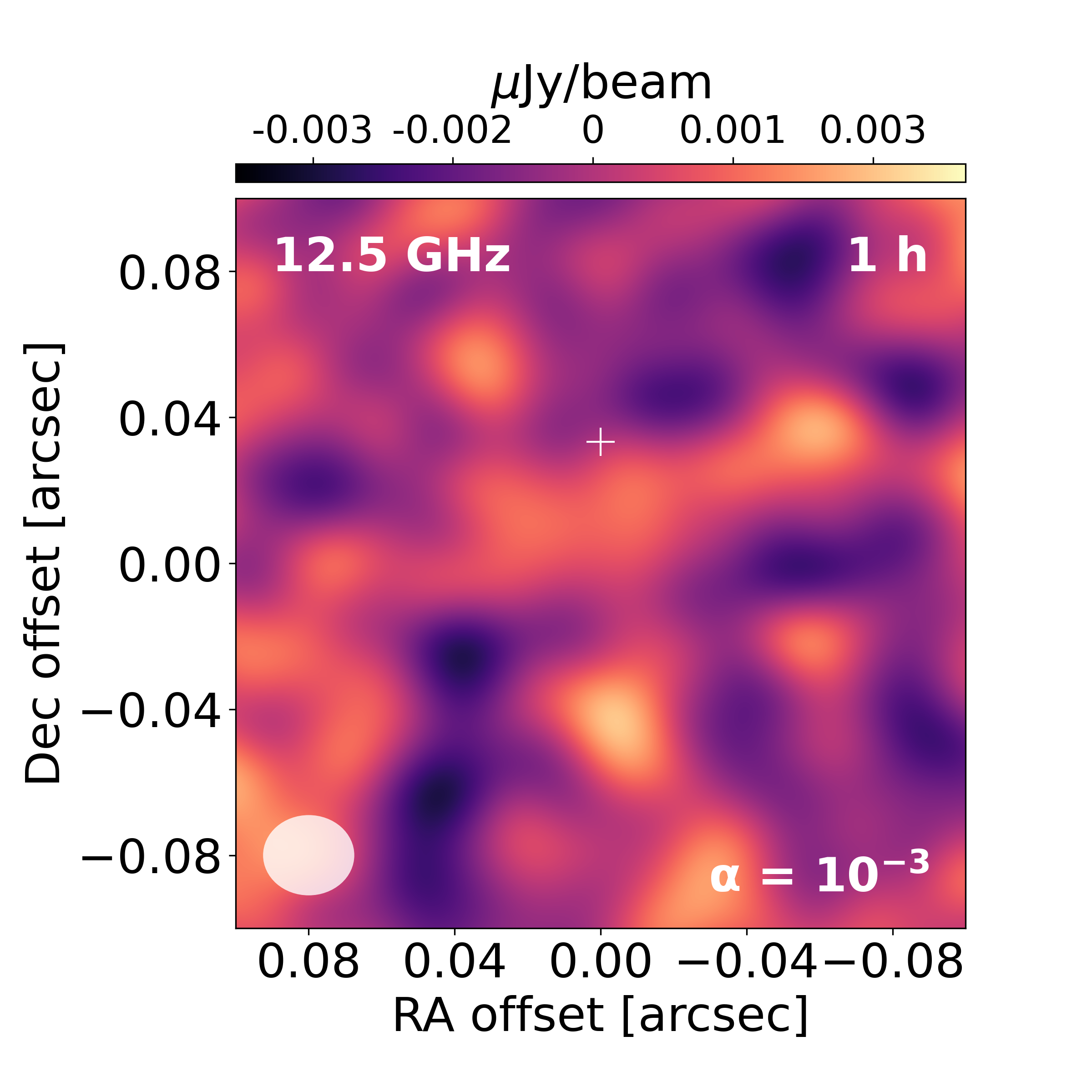}
\includegraphics[width=0.49\hsize]{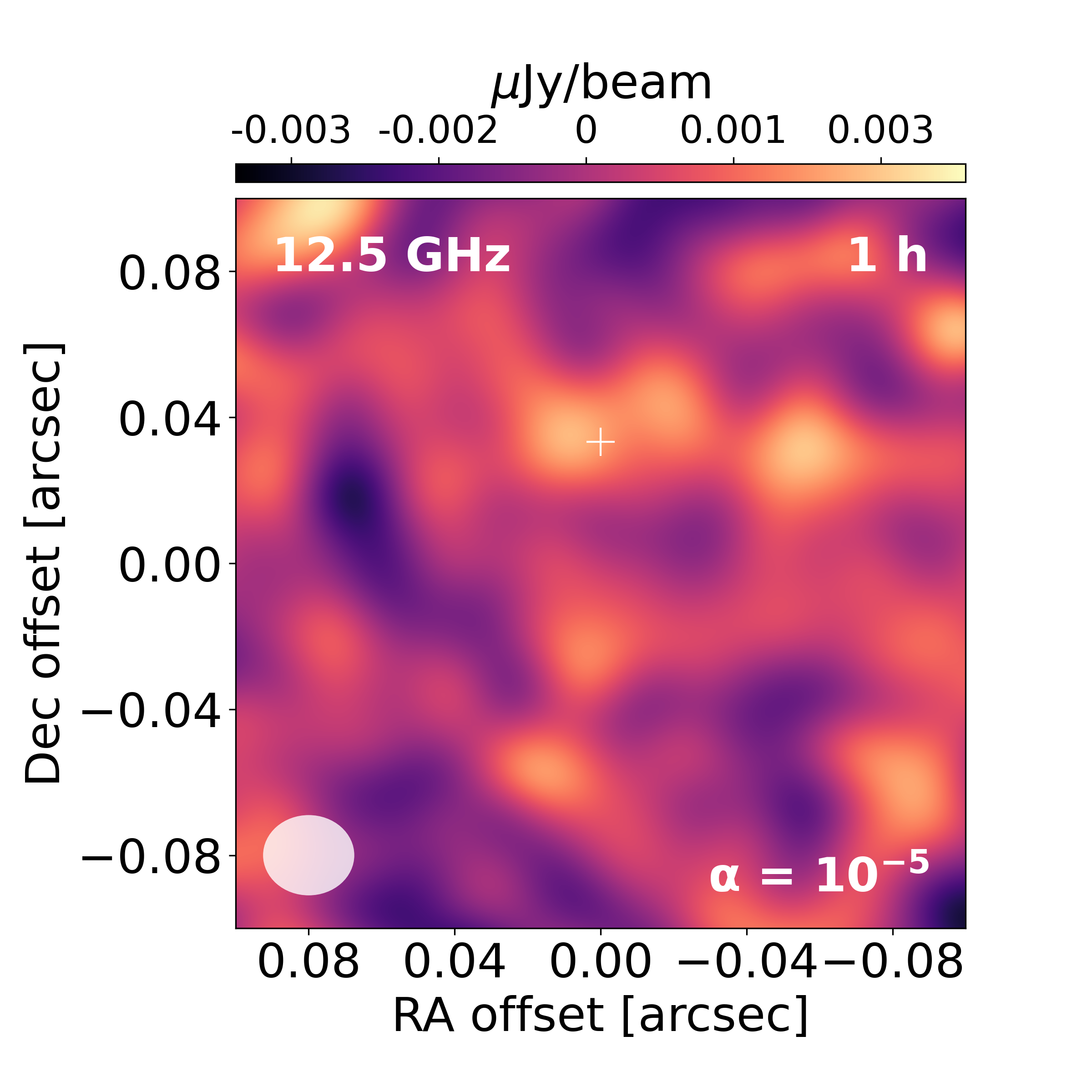}
\includegraphics[width=0.49\hsize]{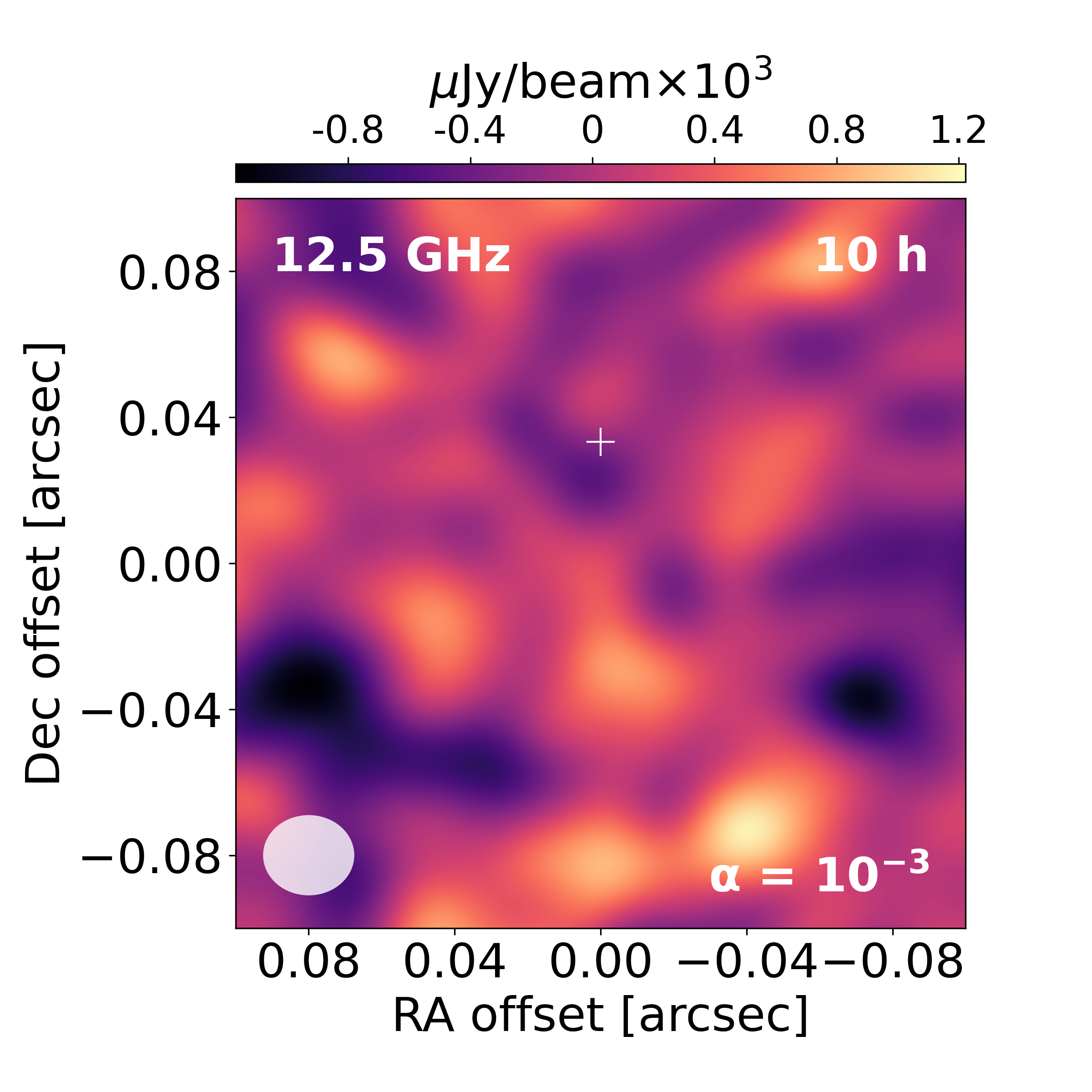}
\includegraphics[width=0.49\hsize]{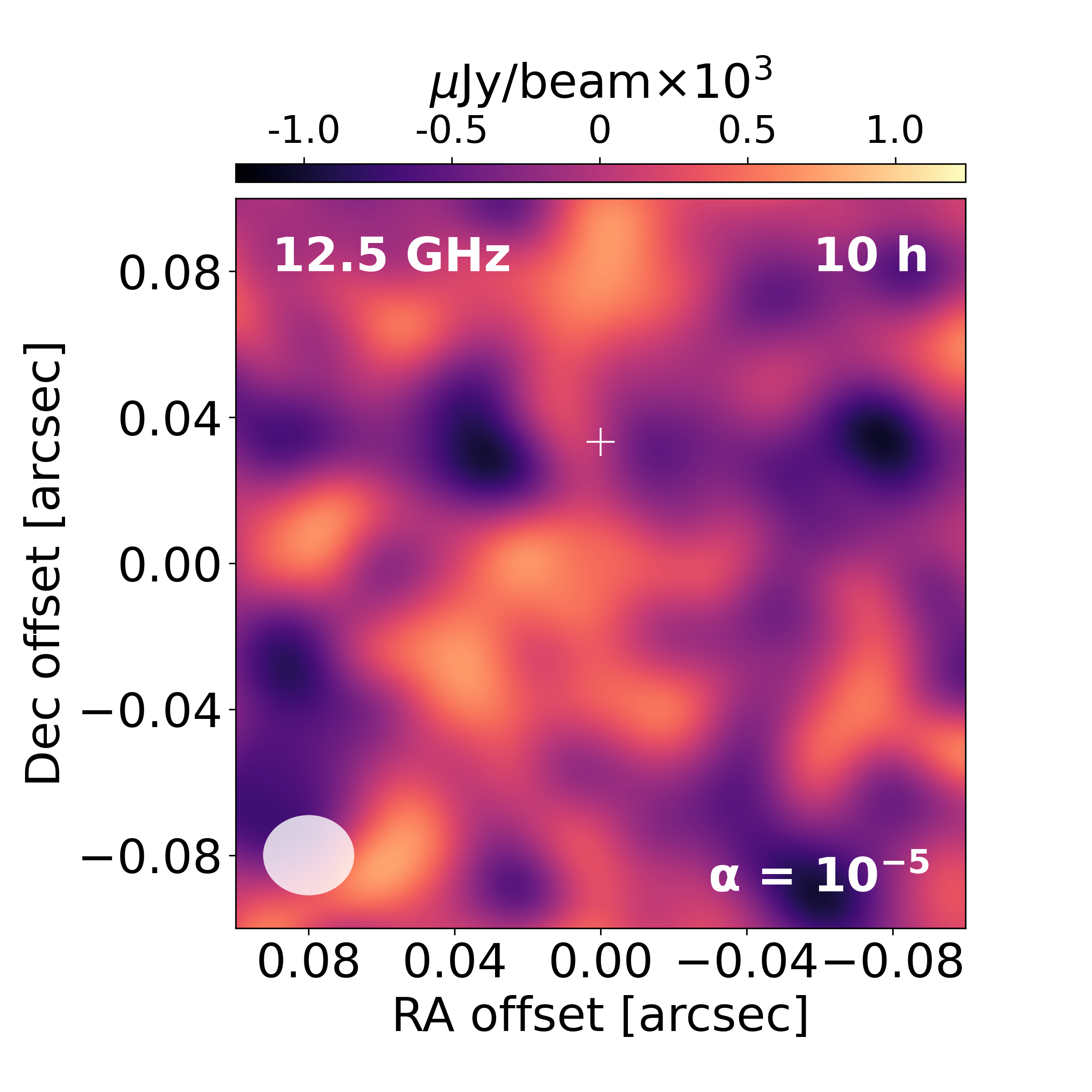}
\includegraphics[width=0.49\hsize]{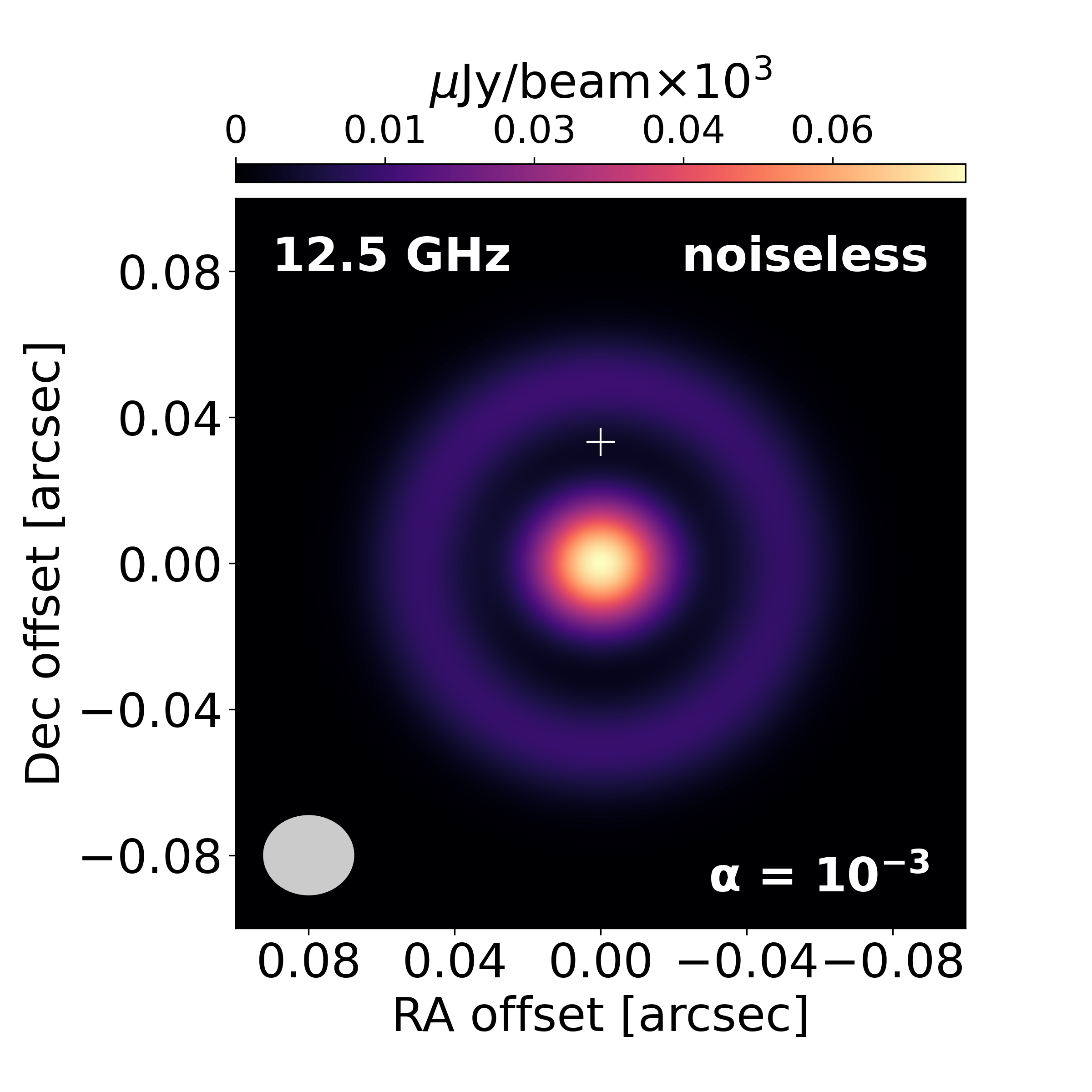}
\includegraphics[width=0.49\hsize]{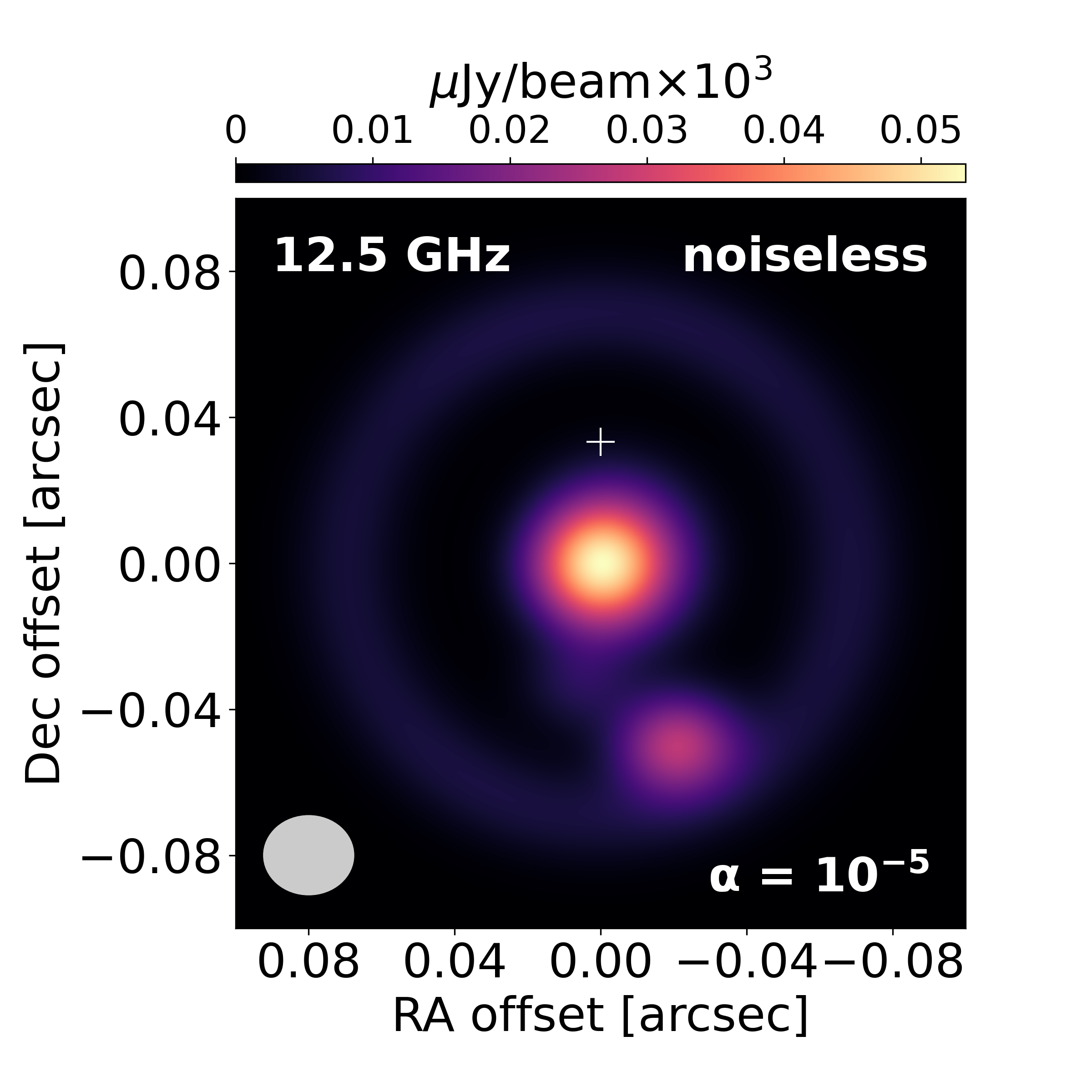}
\caption{Simulated SKA1-Mid (a) continuum images of the planet-disk interaction models with a Jupiter mass planet at 5 au (marked by white ``+") from the central stellar. The band wavelength is 12.5 GHz (marked at upper left corner). The beam size ($0.025\times0.022$ arcsec) is marked at the left-lower corner. From top to bottom, there are images with a noise level equivalent to 1-hour on-source time (marked at upper right corner), 1-hour on-source time and without noise image. From left to right, the $\alpha$ viscosity (marked at right-lower corner) in each model is $10^{-3}$ and $10^{-5}$, respectively.}
\label{fig:image-SKA1 Mid High}
\end{figure}

Fig. \ref{fig:image-SKA1 Mid High} shows the results of the SKA1-Mid (a) observations for models with different $\alpha$ viscosity $\alpha=10^{-3}$ for left panels and $\alpha=10^{-5}$ for right panels). SKA1-Mid (a) represents the high-resolution configuration design for SKA1-Mid, featuring an angular resolution of $0.025\times0.022$ arcsec. As shown in Fig. \ref{fig:opac}, at a wavelength of 2.4 cm, the opacities are predominantly attributed to the scattering contributions from grains of sizes 5 mm, and 1 cm. 

The top panels and the middle panels show the realistic observational outcomes, assuming an observation duration of 1 hour and 10 hours of source time, here the sensitivity is 14.5 $\mu Jy/beam$ (for 1 hour) and 4.6 $\mu Jy/beam$ (for 10 hour). Under this condition, utilizing SKA1-Mid (a) for observations, we are virtually unable to discern any structure within the disk. 

The bottom row displays the mock images without any noise, i.e., the observation results under ideal conditions. Under the resolution and observational wavelengths of SKA1-Mid (a), ideally, we can resolve most substructures appearing in Fig.\ref{fig:hydro}, including the gap at 5 au, the bright ring out the planet, and the vortex in the low-viscosity model. However, SKA1-Mid is incapable of resolving the secondary inner gap within the low-viscosity model, due to the beam size resulting from the resolution exceeding this area (less than 1 au).

\subsection{Observation at 1.87 cm Band with ngVLA}\label{sec:ngVLA 1.87}
\begin{figure}
\centering
\includegraphics[width=0.49\hsize]{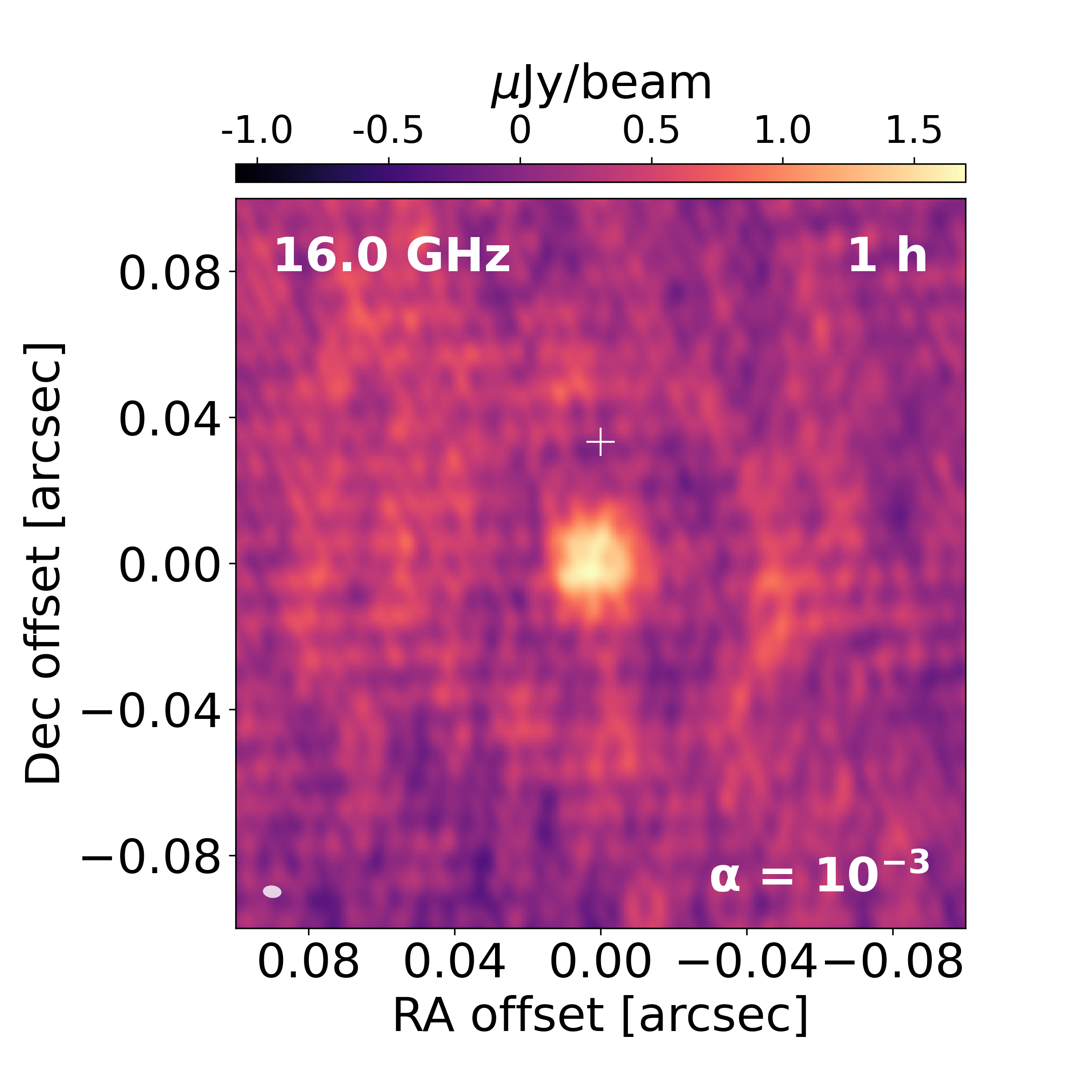}
\includegraphics[width=0.49\hsize]{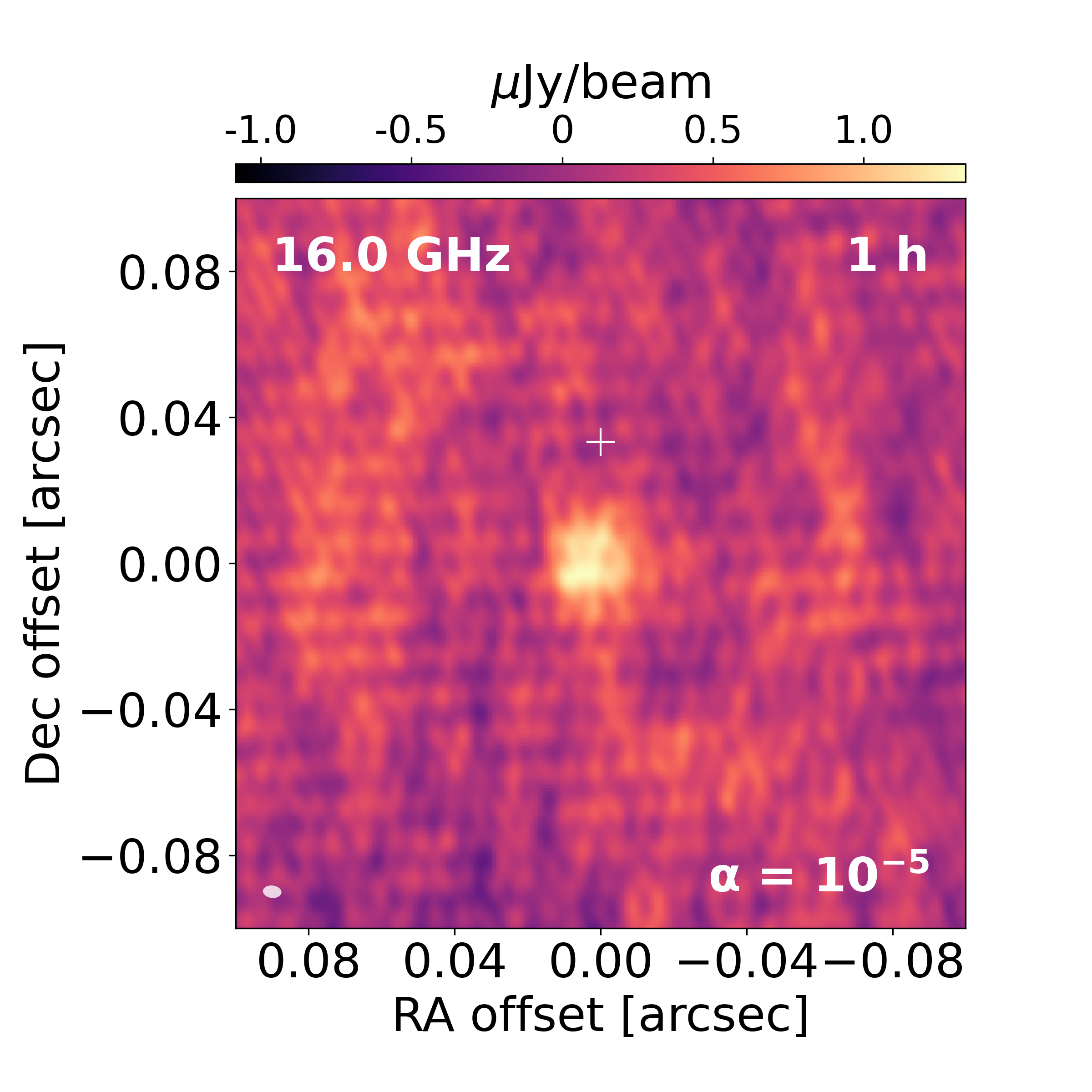}
\includegraphics[width=0.49\hsize]{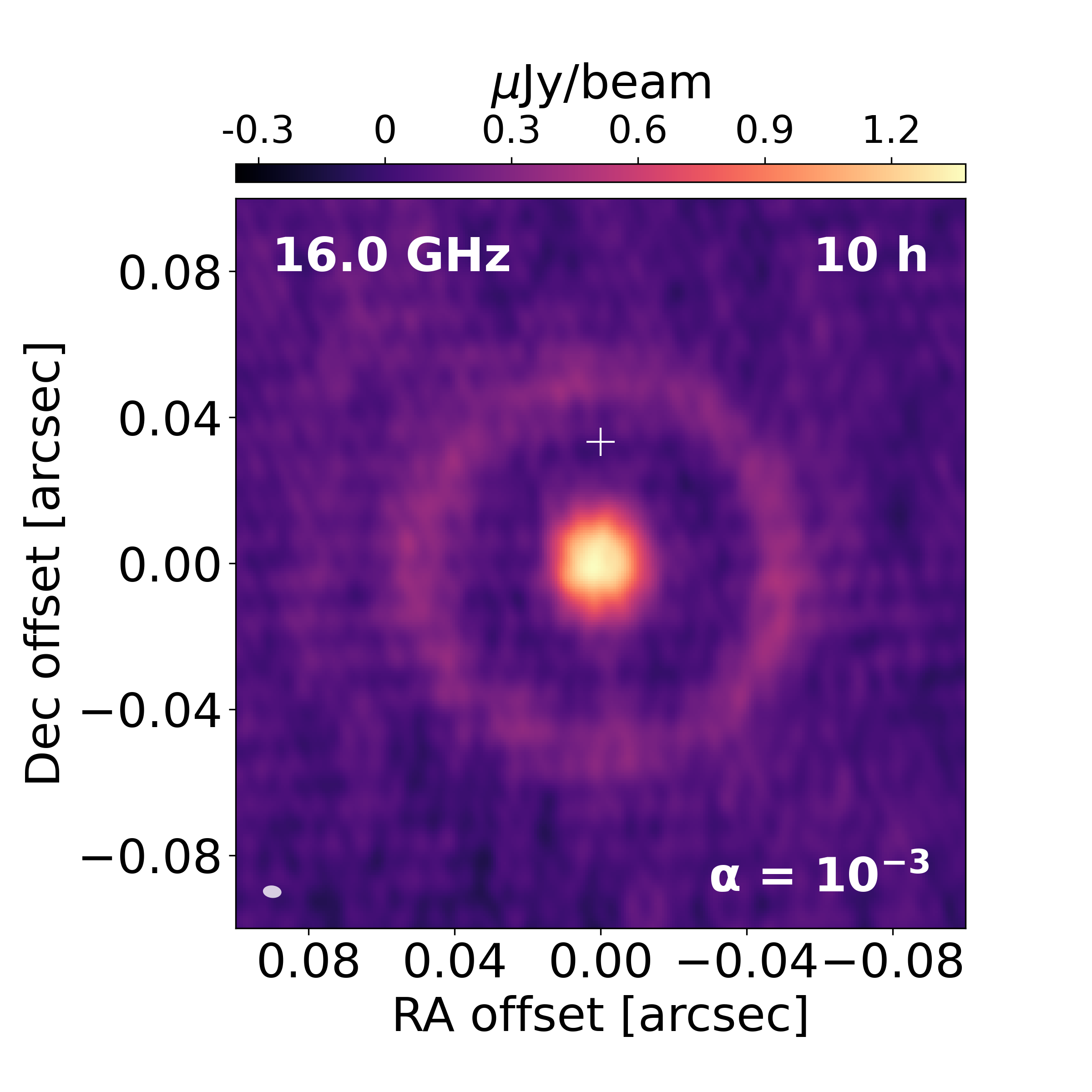}
\includegraphics[width=0.49\hsize]{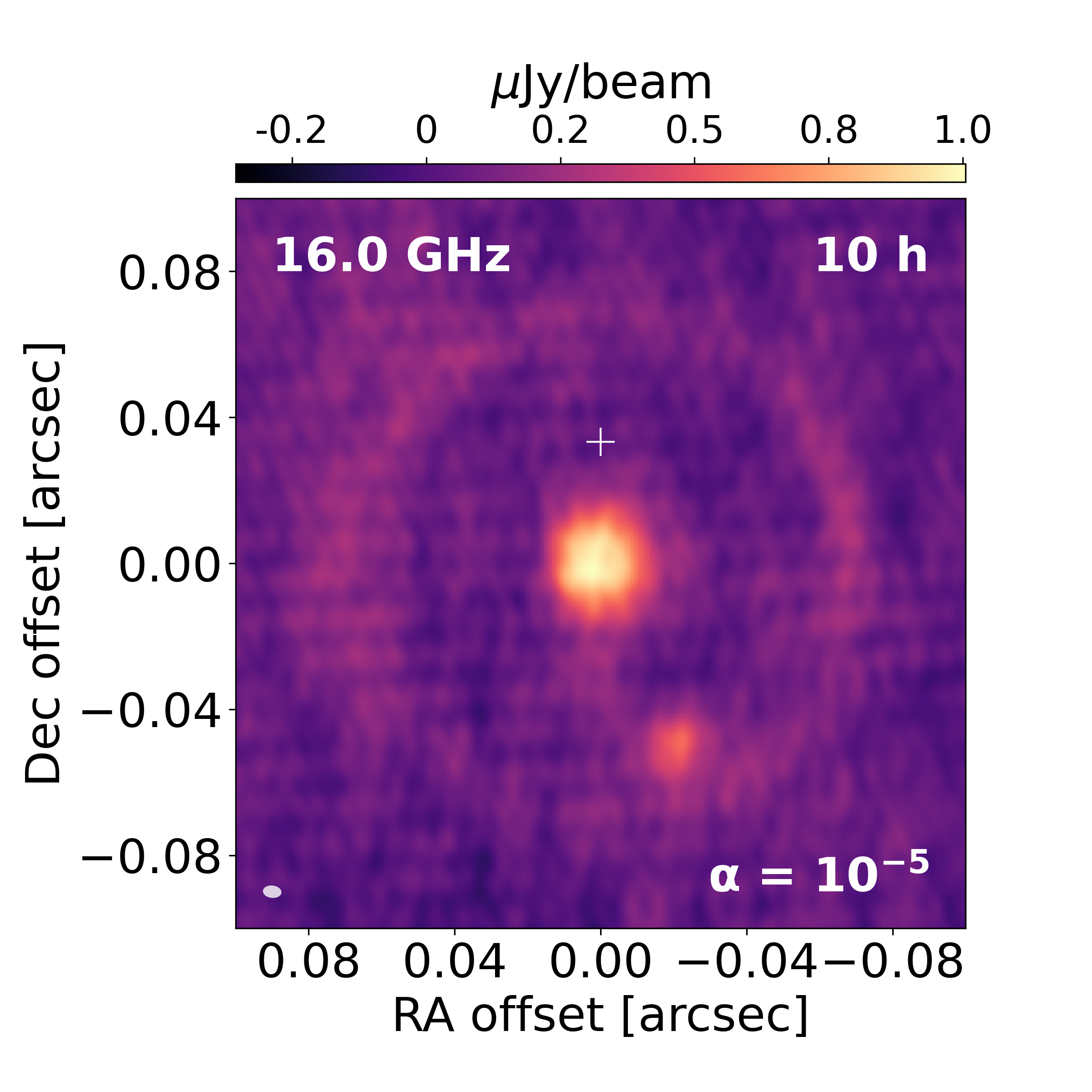}
\caption{Same as 
Fig.\ref{fig:image-SKA1 Mid High}, but for ngVLA Band 3 (naturally weighted point source sensitivity is = 0.23 $\mu Jy/beam$ on 1 hour source time). The wavelength is 1.87 cm (i.e. 16.0 GHz), and beam size is $0.0051\times0.0034$ mas.}
\label{fig:image-ngVLA 1.87}
\end{figure}

For ngVLA Band 3, we evaluated the observational capabilities at a wavelength of 1.87 cm in Fig.\ref{fig:image-ngVLA 1.87}. Referring to Fig.\ref{fig:opac}, we discern that at this wavelength, opacities predominantly arise from the scattering contributions of grains sized at 2.3 mm, 5 mm, and 1 cm. The overall opacities are akin to those observed at a wavelength of 2.4 cm.

In comparison to SKA1-Mid, the ngVLA boasts superior resolution ($0.0051\times0.0034$ arcsec) and sensitivity (0.23 $\mu Jy/beam$ on 1 hour source time). Under long-time observational conditions (such as 10 hours on source time, the bottom panels), ngVLA can resolve most substructures within the disk at a wavelength of 1.87 cm, except the secondary inner gap within the low-viscosity model. However, constrained by the low dust emission rates of cm/sub-cm size dust in the cm-band, ngVLA Band 3 is incapable of resolving the disk's substructures with only 1 hour of source time, except the the emission of stellar.

\subsection{Observation at 1.0 cm Band with ngVLA}\label{sec:ngVLA 1.0}

\begin{figure}
\centering
\includegraphics[width=0.49\hsize]{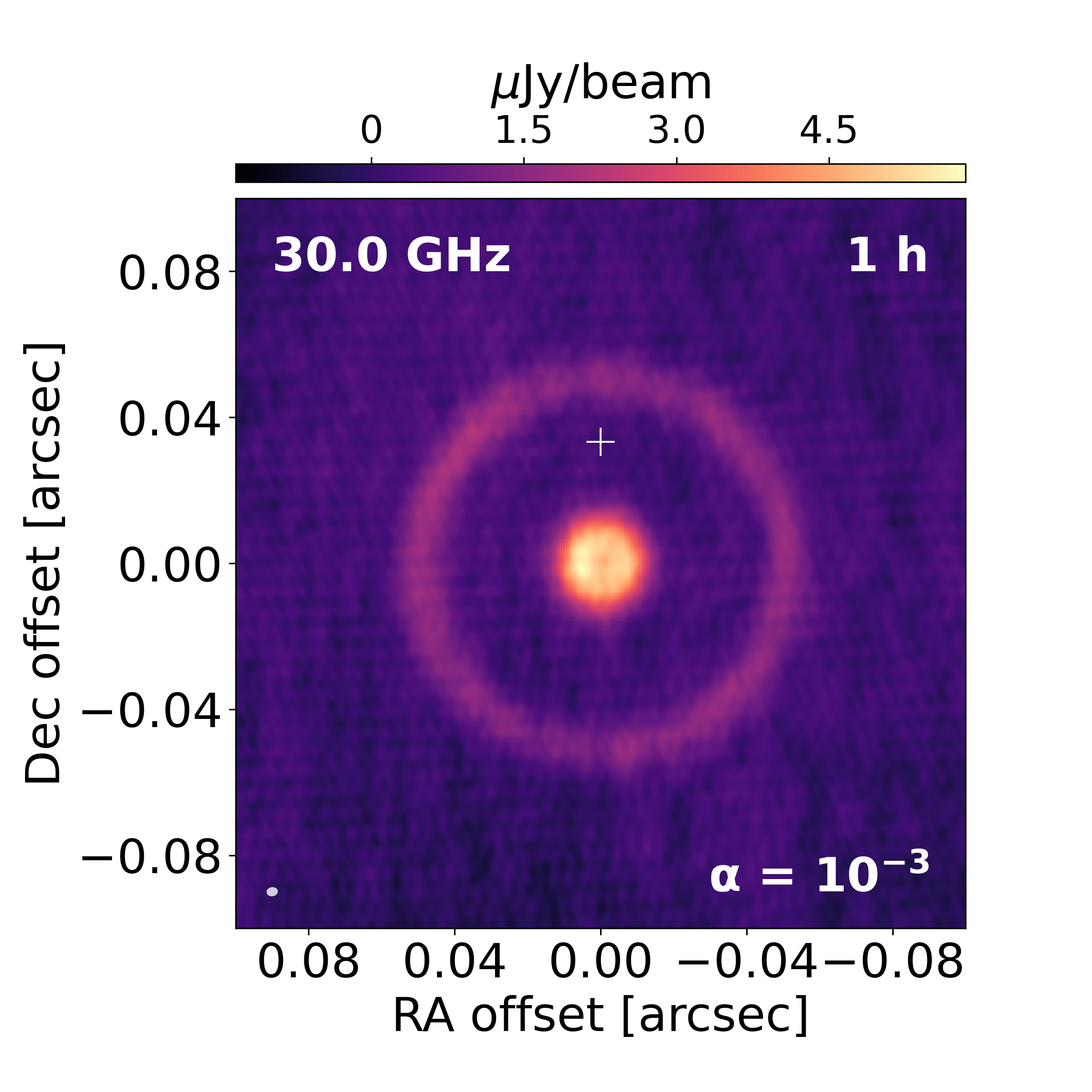}
\includegraphics[width=0.49\hsize]{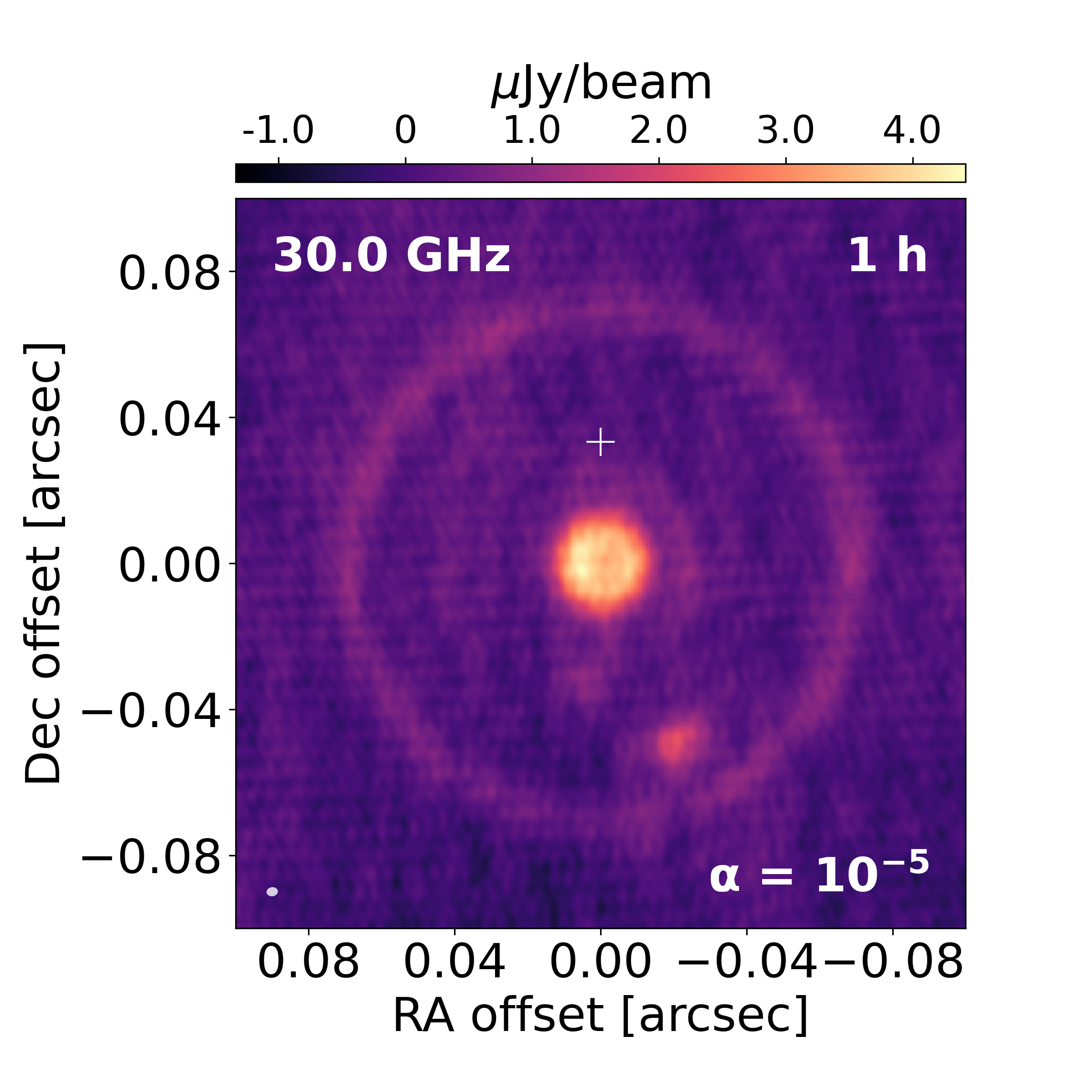}
\caption{Same as 
Fig.\ref{fig:image-SKA1 Mid High}, but for ngVLA Band 4 (naturally weighted point source sensitivity is = 0.24 $\mu Jy/beam$ on 1 hour source time). The wavelength is 1.0 cm (i.e. 30.0 GHz), and beam size is $0.0031\times0.0025$ arcsec.}
\label{fig:image-ngVLA 1.0}
\end{figure}

In Fig.\ref{fig:image-ngVLA 1.0},
we assume naturally weighted point source sensitivity is 0.24 $\mu Jy/beam$ on 1 hour source time for ngVLA observations at the 1 cm wavelength, with an angular resolution of $0.0031\times0.0025$ arcsec, higher than that presented by \cite{Ricci_2018}. We observe that, even with only 1 hour of observation time, ngVLA Band 4 is capable of fully resolving the substructures within the disk at a 1 cm wavelength. Although the signal is quite faint and the signal-to-noise ratio relatively low for the low-viscosity model, it remains significantly above the noise level. At this point, dust emission predominantly originates from grains measuring 1.1 mm, 2.3 mm, 5 mm, and 1 cm.

\subsection{Observation at 0.7 cm Band}\label{sec:sub-cm}
\begin{figure}
\centering
\includegraphics[width=0.49\hsize]{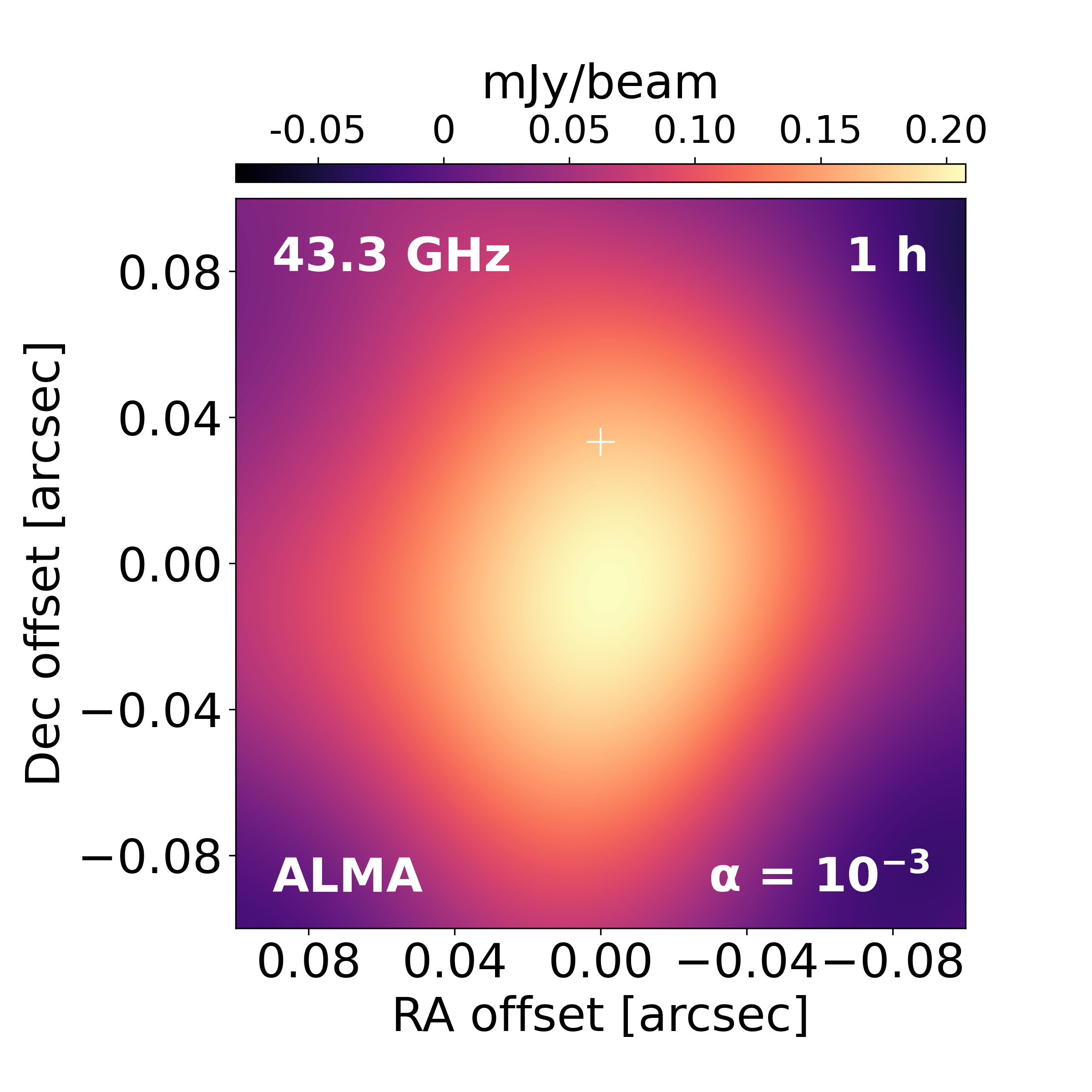}
\includegraphics[width=0.49\hsize]{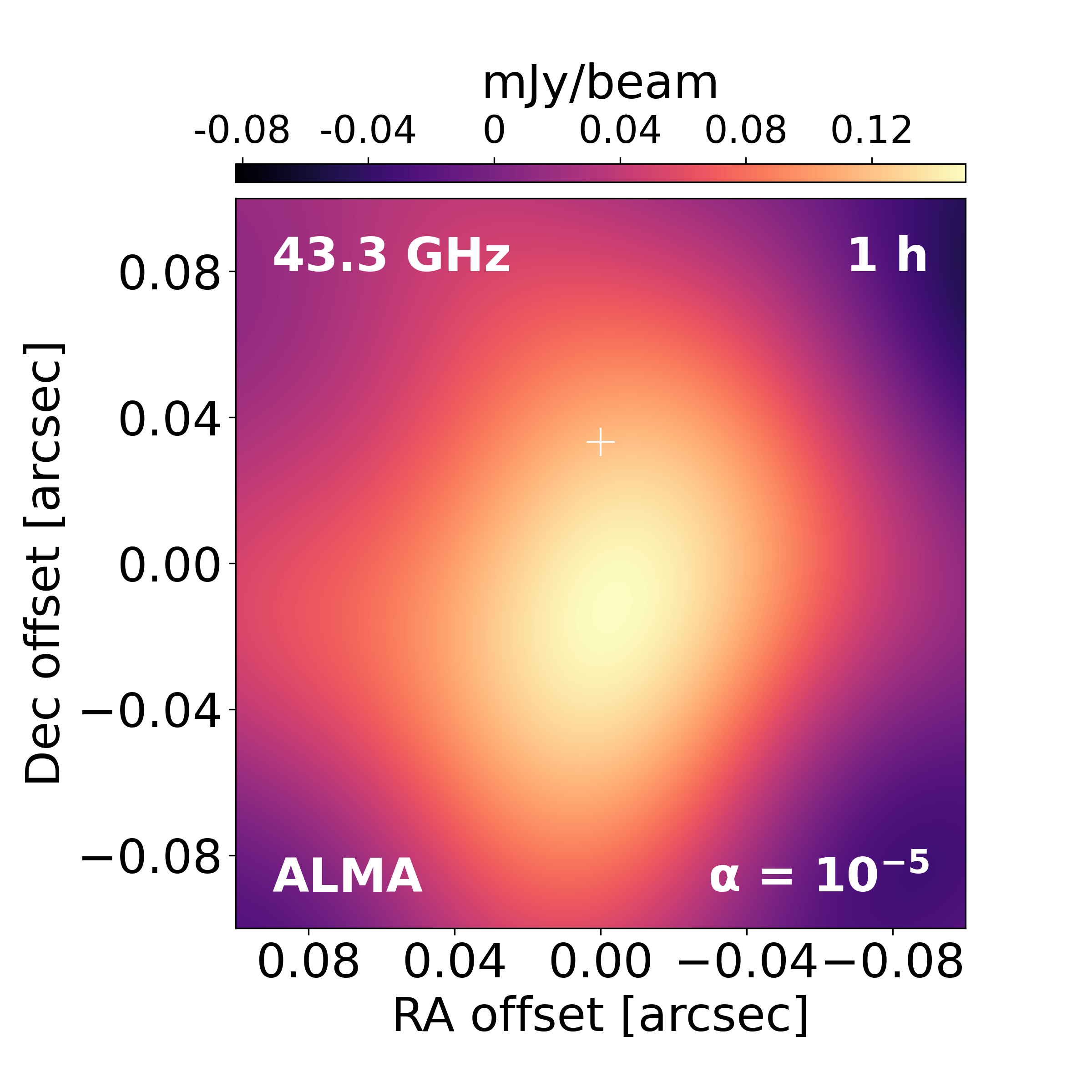}
\includegraphics[width=0.49\hsize]{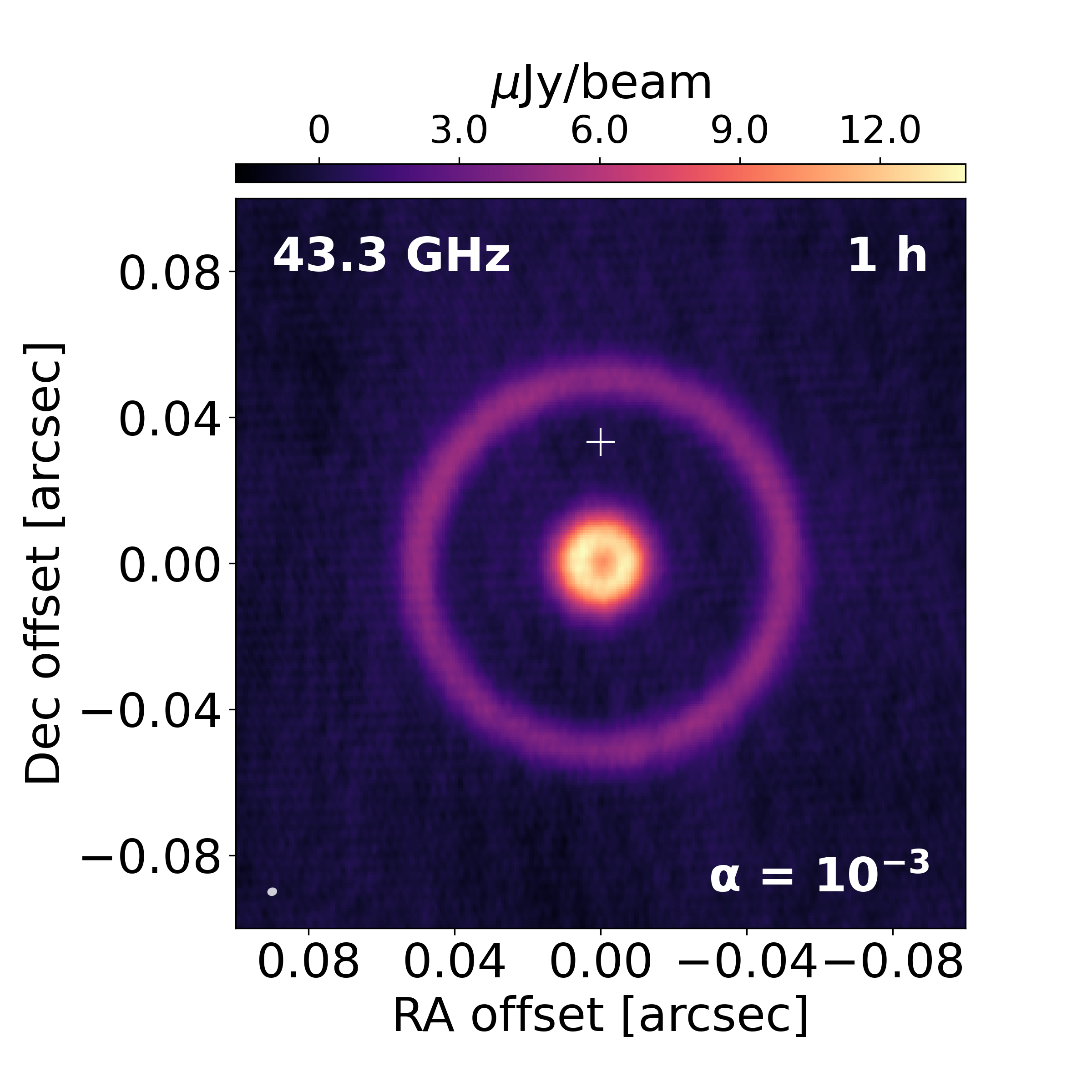}
\includegraphics[width=0.49\hsize]{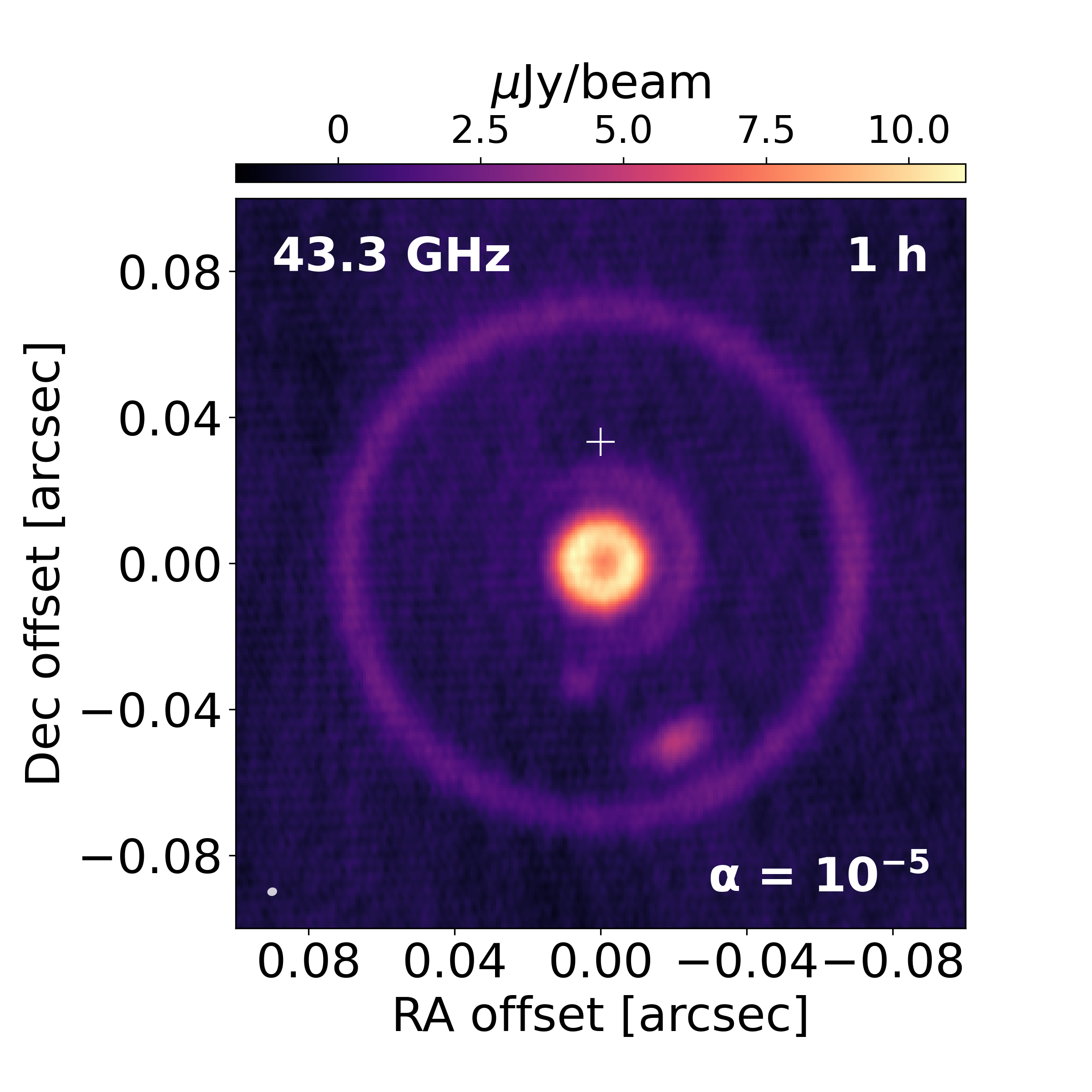}
\includegraphics[width=0.49\hsize]{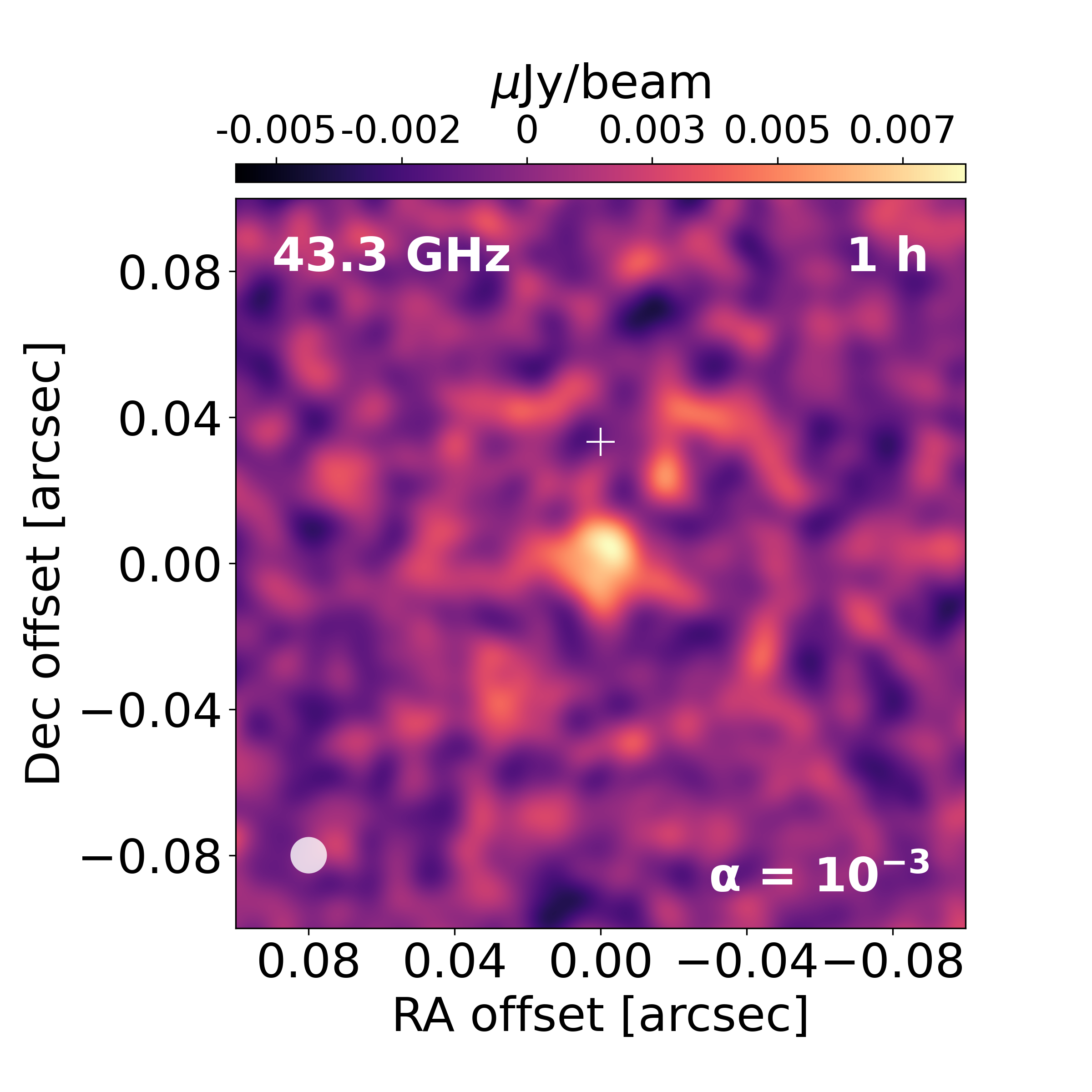}
\includegraphics[width=0.49\hsize]{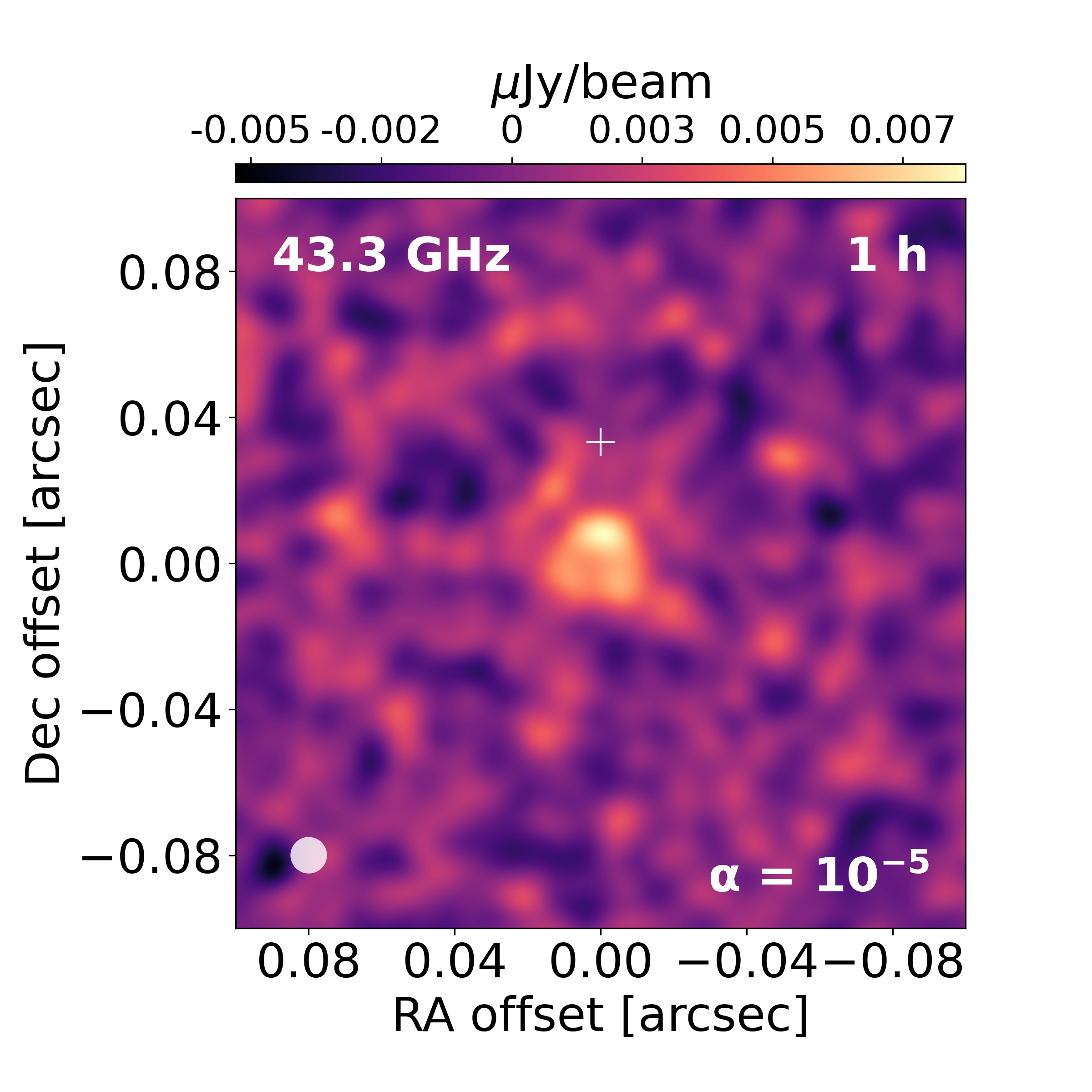}
\includegraphics[width=0.49\hsize]{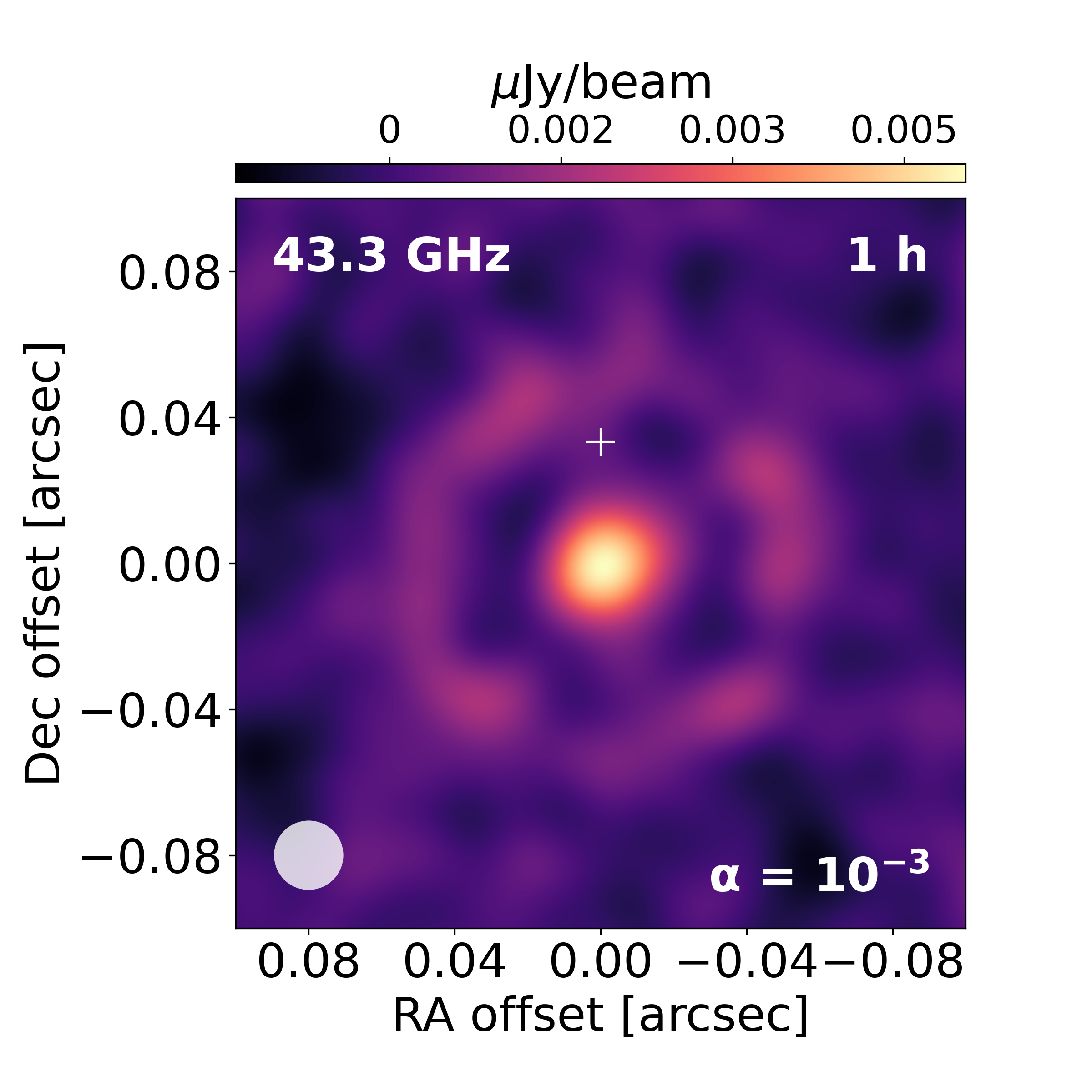}
\includegraphics[width=0.49\hsize]{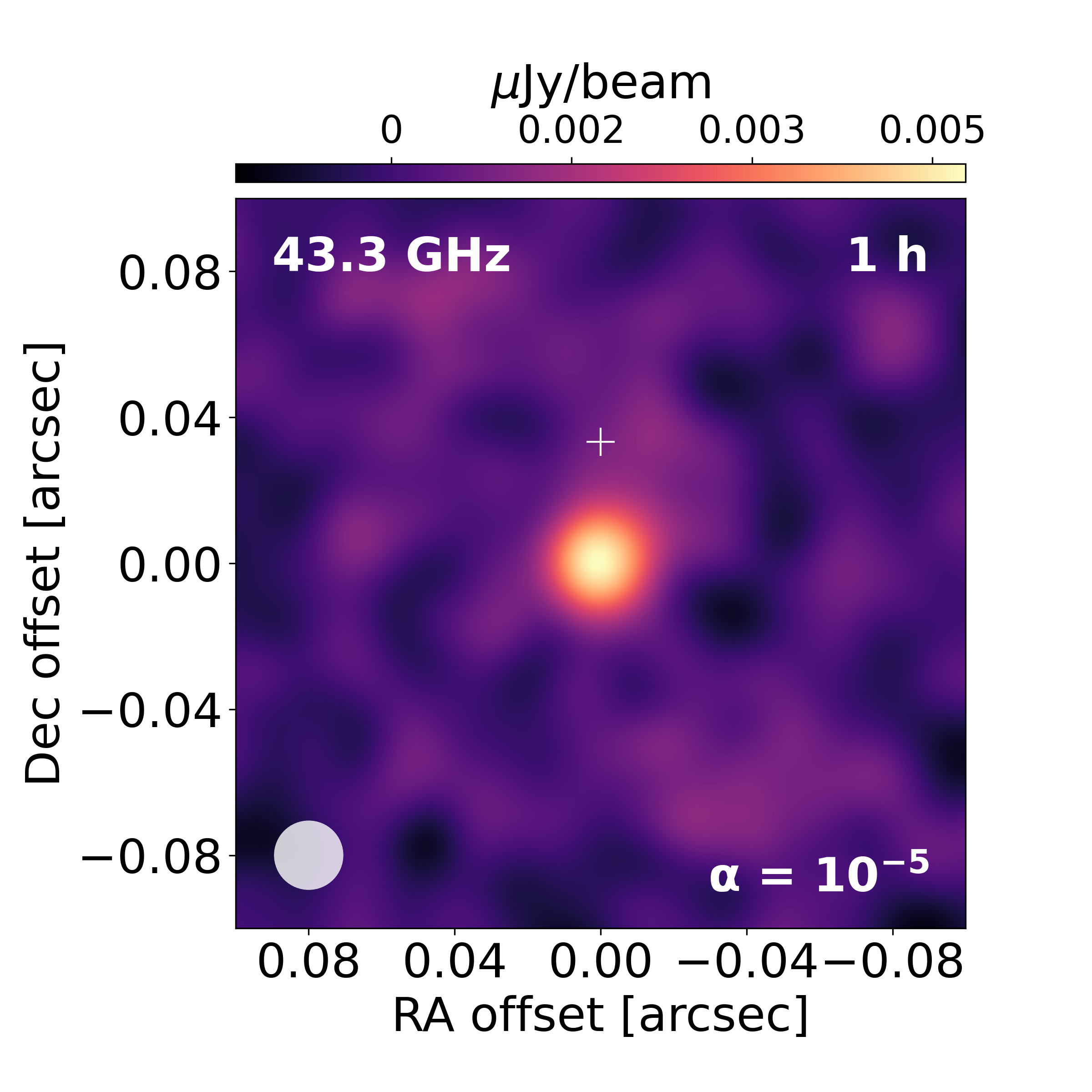}
\caption{Simulated continuum images at 0.7 cm wavelength (i.e. 43.3 GHz) with different facilities. From top to bottom, they are ALMA Band 1 ($0.1\times0.1$ arcsec, rms = 8.5 $\mu Jy/beam$ on 1 hour source time), ngVLA Band 5 ($0.0027\times0.0023$ arcsec,, naturally weighted point source sensitivity is 0.32 $\mu Jy/beam$ on 1 hour source time), SKA2 plan-a ($0.01\times0.01$ arcsec, rms = 3.2 $\mu Jy/beam$ on 1 hour source time), and SKA2 plan-b ($0.019\times0.019$ arcsec,, rms = 3.2 $\mu Jy/beam$ on 1 hour source time). From left to right, same as Fig.\ref{fig:image-SKA1 Mid High}. Due to the excessively large beam size of ALMA Band 1, we did not include the beam in ALMA's panels for convenience.}
\label{fig:image-0.7}
\end{figure}

The designed frequency bands for ngVLA include 43.3 GHz (i.e., $\lambda$ = 0.7 cm, rms = 0.32 $\mu Jy/beam$ on 1 hour source time), which is also among the planned frequency bands for SKA2 (rms = 3.2 $\mu Jy/beam$ on 1 hour source time). Coincidentally, the existing ALMA has just acquired the capability to conduct observations in the sub-cm band, specifically ALMA Band 1, which similarly covers the 0.7 cm wavelength. In Fig.\ref{fig:image-0.7}, we undertake a comparison of the observational capabilities of these three distinct facilities. Within this context, ALMA Band 1 exhibits inferior resolution and sensitivity, standing at 100 mas and rms = 8.5 $\mu Jy/beam$ on 1 hour source time, respectively. SKA2 (plan a) boasts a resolution on par with ngVLA, at $0.01\times0.01$ arcsec Conversely, SKA2 (plan b) presents a slightly reduced resolution of $0.019\times0.019$ arcsec, falling short of both its counterparts.

As illustrated in Fig.\ref{fig:image-0.7}, with an observation time of 1 hour, ALMA Band 1 is incapable of resolving any structures at the scale of the disk, a result that is unsurprising. In contrast, SKA2 is capable of vaguely identifying the general outlines of bright rings and gaps within the high-viscosity model. Interestingly, these substructures exhibit a signal-to-noise ratio in SKA2 (Plan b) that is, despite its lower resolution, even better than the signal-to-noise ratio in the high-resolution SKA2 (Plan a). However, SKA2 falls short in resolving structures within the low-viscosity model. In comparison to these three facilities, ngVLA Band 5 performs exceptionally well, capable of resolving any structures in both models. 

\section{Discussion}\label{sec:4}
The multi-wavelength sim-observation from 0.7 cm to 2.4 cm presented in $\S$\ref{sec:3} shows that the SKA and ngVLA will bridge the observational band-gap left by ALMA in probing substructures, such as gaps and rings, created by the interaction between protoplanetary disks and forming planets. 

We employed a hydrodynamics setup akin to that of \cite{Ricci_2018}, and expanded the observation frequency band of ngVLA, discern a similar observational prowess in the ngVLA to previously anticipated. In contrast to \cite{Ilee_2020}, our approach involved the utilization of a gap-opening planet, thereby testing the SKA's potential in scrutinizing disk substructures. Furthermore, we conducted preliminary studies for the proposed SKA2, anticipating its future contributions to the field.

Moreover, for substructures induced by planets in close proximity to their host stars, such as those near 5 au, both the SKA and ngVLA possess the resolution capacity to delineate them. The ngVLA, in particular, has the potential to resolve vortices and the asymmetry generated by planet-disk interactions, as well as secondary gaps that emerge in more inner regions under conditions of low viscosity. One of the hottest topics of late—the MHD disk wind—is particularly active in the more inner regions of disks \citep[e.g.,][]{Armitage-2013,Bai-2013,Bai-2014,Bai-2016,MHD-wind-Elbakyan,Aoyama-Bai-2023,Wu-Chen-Jiang_2023}. At the same time, compared to broader disk regions, certain effects may have a more significant impact in the inner disk region, such as the Hall effect and strong turbulence, which possess undeniable potential regarding planetesimal formation \citep{Wu_Lin-2024} and planet chaotic migration \citep{Wu-Chen-Lin-2024}. The resolving capabilities of SKA and ngVLA in the inner disk region are instrumental in understanding these effects.

While enhancements in the signal-to-noise ratio for SKA and ngVLA observations at cm wavelengths remain a consideration, sub-cm wavelength observations hold the promise of high-fidelity imaging within reasonable integration times for point observations with these facilities. Consequently, there is an anticipated expansion in the scope of multi-wavelength observations utilizing a broader spectrum range with the SKA and ngVLA in the future.

Recent multi-wavelength studies \citep[e.g.,][]{Carrasco-Gonzalez-2019,Macias-2019,Macias-2021} employing ALMA and the VLA, as well as rare CO isotopologue surveys \citep{Booth-2019,Booth-2020} targeting protoplanetary disks, suggest that disks may indeed be optically thick at submillimeter wavelengths. This condition could lead to misconceptions regarding fundamental characteristics of protoplanetary disks, such as the mass of dust and gas, along with the distribution of dust sizes. By operating at longer wavelengths with the SKA and ngVLA, it might be possible to observe disks that are optically thin, revealing structures that, while present, are less discernible at submillimeter wavelengths due to their concealment beneath optically thick regions \citep{Wu-Chen-Jiang_2023}. The feasibility of this hypothesis has been verified in \cite{Baobab_2019}, as the spectral index of less than 2.0 in the center of TW Hya can only be explained by optically thick dust. Recently, the morphological comparison of DM Tau at 1 mm and 7 mm wavelengths indicates that the extremely optically thick at 1 mm \citep{Baobab_2024}.

\section{Conclusions}\label{sec:5}
We present the hydrodynamic simulation results from \texttt{FARGO3D}, encompassing disks with embedded planets and the intricate interactions between disk gas and dust. The primary aim of this study is to explore the capabilities of current (ALMA Band 1) and forthcoming (SKA and ngVLA) telescopic facilities in probing planet formation at sub-cm/cm wavelengths. Our main findings are as follows:

\begin{itemize}
    \item SKA1-Mid, ngVLA, and SKA2 all demonstrate the capacity to resolve substructures in protoplanetary disks as well as their inner regions at sub-cm/cm wavelengths.

    \item At cm wavelengths, SKA1-Mid and ngVLA Band 3 require extended integration times and high resolution. For studies of protoplanetary disks, future enhancements should prioritize boosting the signal-to-noise ratio to facilitate research within more reasonable observational timeframes. Alternatively, emphasis could be placed on more extended monitoring programs.
    
    \item The ngVLA exhibits exceptional performance in the sub-centimeter bands (Band 4 and Band 5), capable of producing high-fidelity images of disk substructures due to gap-opening planets within feasible integration times (less than 1 hr on source time). It holds promise for future high-resolution surveys.
    
    \item While SKA2 shares a similar observational band with ngVLA Band 5, its ability to detect planet signatures appears somewhat less strong in comparison. However, it will serve as a valuable complement to SKA1-Mid.
\end{itemize}

In summary, both SKA and ngVLA will serve as significant complements to the existing (sub)mm facility, e.g., ALMA. Our research indicates that SKA and ngVLA will play a crucial role in characterizing the distribution and properties of pebbles within protoplanetary disks. In subsequent studies, we plan to further explore how to employ SKA and ngVLA in conjunction with ALMA for multi-wavelength collaborative observations using SED methods, aiming to better constrain the distribution of dust sizes.



\section*{Aknowledgement}
We thank our referee for highly constructive suggestions, which significantly enhanced the quality of the manuscript. We thank Ruobing Dong for his help and useful discussions in the project. We thank Cl\'ement Baruteau and Pablo Beni\'tez-Llambay for their helps on multi-fluid code development suggestions. We thank Yixian Chen, Pinghui Huang and Yaping Li on the analysis of the preliminary hydrodynamic results. We thank Shangjia Zhang for his help on calculation of radiation transfer. We thank John Ilee and Fuguo Xie for their discussions on SKA. We thank Yihuan Di, Tie Liu, Feng Long, Qiuyi Luo and Giovanni Rosotti for their assistance with the use of CASA. S.-F.L. acknowledges support from National Natural Science Foundation of China under grant No. 11903089 and from Guangdong Basic and Applied Basic Research Foundation under grant Nos. 2021B1515020090 and 2019B030302001, and from the China Manned Space Project under grant Nos. CMS-CSST-2021-B02 and CMS-CSST-2021-B12.

This work was in part supported by the National SKA Program of China. This research used TianHe-2, a supercomputer owned by the National Supercomputer Center in Guangzhou, China, and the Kunlun cluster at School of Physics and Astronomy, Sun Yat-sen University. Part of the numerical simulations were performed on the ALICE High Performance Computing Facility at the University of Leicester, and DiRAC Data Intensive service at Leicester, operated by the University of Leicester IT Services, which forms part of the STFC DiRAC HPC Facility (\href{www.dirac.ac.uk}{www.dirac.ac.uk}).

\section*{Data availability}
The data obtained in our simulations can be made available on reasonable request to the corresponding author. 

\bibliography{main-body}{}

\begin{thebibliography}{}
\expandafter\ifx\csname natexlab\endcsname\relax\def\natexlab#1{#1}\fi
\providecommand{\url}[1]{\href{#1}{#1}}
\providecommand{\dodoi}[1]{doi:~\href{http://doi.org/#1}{\nolinkurl{#1}}}
\providecommand{\doeprint}[1]{\href{http://ascl.net/#1}{\nolinkurl{http://ascl.net/#1}}}
\providecommand{\doarXiv}[1]{\href{https://arxiv.org/abs/#1}{\nolinkurl{https://arxiv.org/abs/#1}}}

\bibitem[{{ALMA Partnership} {et~al.}(2015){ALMA Partnership}, {Brogan}, {P{\'e}rez}, {Hunter}, {Dent}, {Hales}, {Hills}, {Corder}, {Fomalont}, {Vlahakis}, {Asaki}, {Barkats}, {Hirota}, {Hodge}, {Impellizzeri}, {Kneissl}, {Liuzzo}, {Lucas}, {Marcelino}, {Matsushita}, {Nakanishi}, {Phillips}, {Richards}, {Toledo}, {Aladro}, {Broguiere}, {Cortes}, {Cortes}, {Espada}, {Galarza}, {Garcia-Appadoo}, {Guzman-Ramirez}, {Humphreys}, {Jung}, {Kameno}, {Laing}, {Leon}, {Marconi}, {Mignano}, {Nikolic}, {Nyman}, {Radiszcz}, {Remijan}, {Rod{\'o}n}, {Sawada}, {Takahashi}, {Tilanus}, {Vila Vilaro}, {Watson}, {Wiklind}, {Akiyama}, {Chapillon}, {de Gregorio-Monsalvo}, {Di Francesco}, {Gueth}, {Kawamura}, {Lee}, {Nguyen Luong}, {Mangum}, {Pietu}, {Sanhueza}, {Saigo}, {Takakuwa}, {Ubach}, {van Kempen}, {Wootten}, {Castro-Carrizo}, {Francke}, {Gallardo}, {Garcia}, {Gonzalez}, {Hill}, {Kaminski}, {Kurono}, {Liu}, {Lopez}, {Morales}, {Plarre}, {Schieven}, {Testi}, {Videla}, {Villard}, {Andreani}, {Hibbard}, \&
  {Tatematsu}}]{ALMA2015}
{ALMA Partnership}, {Brogan}, C.~L., {P{\'e}rez}, L.~M., {et~al.} 2015, \apjl, 808, L3, \dodoi{10.1088/2041-8205/808/1/L3}

\bibitem[{{Andrews}(2020)}]{Andrews-araa-2020}
{Andrews}, S.~M. 2020, \araa, 58, 483, \dodoi{10.1146/annurev-astro-031220-010302}

\bibitem[{{Andrews} {et~al.}(2018){Andrews}, {Huang}, {P{\'e}rez}, {Isella}, {Dullemond}, {Kurtovic}, {Guzm{\'a}n}, {Carpenter}, {Wilner}, {Zhang}, {Zhu}, {Birnstiel}, {Bai}, {Benisty}, {Hughes}, {{\"O}berg}, \& {Ricci}}]{DSHARP-I}
{Andrews}, S.~M., {Huang}, J., {P{\'e}rez}, L.~M., {et~al.} 2018, \apjl, 869, L41, \dodoi{10.3847/2041-8213/aaf741}

\bibitem[{{Aoyama} \& {Bai}(2023)}]{Aoyama-Bai-2023}
{Aoyama}, Y., \& {Bai}, X.-N. 2023, \apj, 946, 5, \dodoi{10.3847/1538-4357/acb81f}

\bibitem[{{Armitage} {et~al.}(2013){Armitage}, {Simon}, \& {Martin}}]{Armitage-2013}
{Armitage}, P.~J., {Simon}, J.~B., \& {Martin}, R.~G. 2013, \apjl, 778, L14, \dodoi{10.1088/2041-8205/778/1/L14}

\bibitem[{{Bai} \& {Stone}(2013)}]{Bai-2013}
{Bai}, X.-N., \& {Stone}, J.~M. 2013, \apj, 769, 76, \dodoi{10.1088/0004-637X/769/1/76}

\bibitem[{{Bai} \& {Stone}(2014)}]{Bai-2014}
---. 2014, \apj, 796, 31, \dodoi{10.1088/0004-637X/796/1/31}

\bibitem[{{Bai} {et~al.}(2016){Bai}, {Ye}, {Goodman}, \& {Yuan}}]{Bai-2016}
{Bai}, X.-N., {Ye}, J., {Goodman}, J., \& {Yuan}, F. 2016, \apj, 818, 152, \dodoi{10.3847/0004-637X/818/2/152}

\bibitem[{{Baruteau} {et~al.}(2021){Baruteau}, {Wafflard-Fernandez}, {Le Gal}, {Debras}, {Carmona}, {Fuente}, \& {Rivi{\`e}re-Marichalar}}]{Baruteau21}
{Baruteau}, C., {Wafflard-Fernandez}, G., {Le Gal}, R., {et~al.} 2021, \mnras, 505, 359, \dodoi{10.1093/mnras/stab1045}

\bibitem[{{Baruteau} {et~al.}(2014){Baruteau}, {Crida}, {Paardekooper}, {Masset}, {Guilet}, {Bitsch}, {Nelson}, {Kley}, \& {Papaloizou}}]{Baruteau-PPVI}
{Baruteau}, C., {Crida}, A., {Paardekooper}, S.~J., {et~al.} 2014, in Protostars and Planets VI, ed. H.~{Beuther}, R.~S. {Klessen}, C.~P. {Dullemond}, \& T.~{Henning}, 667--689, \dodoi{10.2458/azu_uapress_9780816531240-ch029}

\bibitem[{{Baruteau} {et~al.}(2019){Baruteau}, {Barraza}, {P{\'e}rez}, {Casassus}, {Dong}, {Lyra}, {Marino}, {Christiaens}, {Zhu}, {Carmona}, {Debras}, \& {Alarcon}}]{fargo2radmc3d-dust}
{Baruteau}, C., {Barraza}, M., {P{\'e}rez}, S., {et~al.} 2019, \mnras, 486, 304, \dodoi{10.1093/mnras/stz802}

\bibitem[{{Ben{\'\i}tez-Llambay} {et~al.}(2019){Ben{\'\i}tez-Llambay}, {Krapp}, \& {Pessah}}]{FARGO3D-multifluid}
{Ben{\'\i}tez-Llambay}, P., {Krapp}, L., \& {Pessah}, M.~E. 2019, \apjs, 241, 25, \dodoi{10.3847/1538-4365/ab0a0e}

\bibitem[{{Ben{\'\i}tez-Llambay} \& {Masset}(2016)}]{FARGO3D}
{Ben{\'\i}tez-Llambay}, P., \& {Masset}, F.~S. 2016, \apjs, 223, 11, \dodoi{10.3847/0067-0049/223/1/11}

\bibitem[{{Birnstiel} {et~al.}(2018){Birnstiel}, {Dullemond}, {Zhu}, {Andrews}, {Bai}, {Wilner}, {Carpenter}, {Huang}, {Isella}, {Benisty}, {P{\'e}rez}, \& {Zhang}}]{DSHARP-V}
{Birnstiel}, T., {Dullemond}, C.~P., {Zhu}, Z., {et~al.} 2018, \apjl, 869, L45, \dodoi{10.3847/2041-8213/aaf743}

\bibitem[{{Blum} \& {Wurm}(2008)}]{Blum-2008}
{Blum}, J., \& {Wurm}, G. 2008, \araa, 46, 21, \dodoi{10.1146/annurev.astro.46.060407.145152}

\bibitem[{{Booth} \& {Ilee}(2020)}]{Booth-2020}
{Booth}, A.~S., \& {Ilee}, J.~D. 2020, \mnras, 493, L108, \dodoi{10.1093/mnrasl/slaa014}

\bibitem[{{Booth} {et~al.}(2019){Booth}, {Walsh}, {Ilee}, {Notsu}, {Qi}, {Nomura}, \& {Akiyama}}]{Booth-2019}
{Booth}, A.~S., {Walsh}, C., {Ilee}, J.~D., {et~al.} 2019, \apjl, 882, L31, \dodoi{10.3847/2041-8213/ab3645}

\bibitem[{{Braun} {et~al.}(2019){Braun}, {Bonaldi}, {Bourke}, {Keane}, \& {Wagg}}]{Braun-2019-SKA}
{Braun}, R., {Bonaldi}, A., {Bourke}, T., {Keane}, E., \& {Wagg}, J. 2019, arXiv e-prints, arXiv:1912.12699, \dodoi{10.48550/arXiv.1912.12699}

\bibitem[{{Braun} {et~al.}(2015){Braun}, {Bourke}, {Green}, {Keane}, \& {Wagg}}]{Braun-2015-SKA}
{Braun}, R., {Bourke}, T., {Green}, J.~A., {Keane}, E., \& {Wagg}, J. 2015, in Advancing Astrophysics with the Square Kilometre Array (AASKA14), 174, \dodoi{10.22323/1.215.0174}

\bibitem[{{Carrasco-Gonz{\'a}lez} {et~al.}(2019){Carrasco-Gonz{\'a}lez}, {Sierra}, {Flock}, {Zhu}, {Henning}, {Chandler}, {Galv{\'a}n-Madrid}, {Mac{\'\i}as}, {Anglada}, {Linz}, {Osorio}, {Rodr{\'\i}guez}, {Testi}, {Torrelles}, {P{\'e}rez}, \& {Liu}}]{Carrasco-Gonzalez-2019}
{Carrasco-Gonz{\'a}lez}, C., {Sierra}, A., {Flock}, M., {et~al.} 2019, \apj, 883, 71, \dodoi{10.3847/1538-4357/ab3d33}

\bibitem[{{Clarke} {et~al.}(2018){Clarke}, {Tazzari}, {Juhasz}, {Rosotti}, {Booth}, {Facchini}, {Ilee}, {Johns-Krull}, {Kama}, {Meru}, \& {Prato}}]{Clarke-2018}
{Clarke}, C.~J., {Tazzari}, M., {Juhasz}, A., {et~al.} 2018, \apjl, 866, L6, \dodoi{10.3847/2041-8213/aae36b}

\bibitem[{{Dipierro} {et~al.}(2018){Dipierro}, {Ricci}, {P{\'e}rez}, {Lodato}, {Alexander}, {Laibe}, {Andrews}, {Carpenter}, {Chandler}, {Greaves}, {Hall}, {Henning}, {Kwon}, {Linz}, {Mundy}, {Sargent}, {Tazzari}, {Testi}, \& {Wilner}}]{Dipierro-2018}
{Dipierro}, G., {Ricci}, L., {P{\'e}rez}, L., {et~al.} 2018, \mnras, 475, 5296, \dodoi{10.1093/mnras/sty181}

\bibitem[{{Dominik} {et~al.}(2021){Dominik}, {Min}, \& {Tazaki}}]{optool-2021}
{Dominik}, C., {Min}, M., \& {Tazaki}, R. 2021, {OpTool: Command-line driven tool for creating complex dust opacities}, Astrophysics Source Code Library, record ascl:2104.010.
\newblock \doeprint{2104.010}

\bibitem[{{Dong} {et~al.}(2017){Dong}, {Li}, {Chiang}, \& {Li}}]{Dong2017}
{Dong}, R., {Li}, S., {Chiang}, E., \& {Li}, H. 2017, \apj, 843, 127, \dodoi{10.3847/1538-4357/aa72f2}

\bibitem[{{Draine}(2006)}]{Draine-2006}
{Draine}, B.~T. 2006, \apj, 636, 1114, \dodoi{10.1086/498130}

\bibitem[{{Draine} \& {Lee}(1984)}]{Draine-1984}
{Draine}, B.~T., \& {Lee}, H.~M. 1984, \apj, 285, 89, \dodoi{10.1086/162480}

\bibitem[{{Dullemond} {et~al.}(2012){Dullemond}, {Juhasz}, {Pohl}, {Sereshti}, {Shetty}, {Peters}, {Commercon}, \& {Flock}}]{RADMC-3D-2012}
{Dullemond}, C.~P., {Juhasz}, A., {Pohl}, A., {et~al.} 2012, {RADMC-3D: A multi-purpose radiative transfer tool}, Astrophysics Source Code Library, record ascl:1202.015.
\newblock \doeprint{1202.015}

\bibitem[{{Elbakyan} {et~al.}(2022){Elbakyan}, {Wu}, {Nayakshin}, \& {Rosotti}}]{MHD-wind-Elbakyan}
{Elbakyan}, V., {Wu}, Y., {Nayakshin}, S., \& {Rosotti}, G. 2022, \mnras, 515, 3113, \dodoi{10.1093/mnras/stac1774}

\bibitem[{{Fedele} {et~al.}(2017){Fedele}, {Carney}, {Hogerheijde}, {Walsh}, {Miotello}, {Klaassen}, {Bruderer}, {Henning}, \& {van Dishoeck}}]{HD169142-2017}
{Fedele}, D., {Carney}, M., {Hogerheijde}, M.~R., {et~al.} 2017, \aap, 600, A72, \dodoi{10.1051/0004-6361/201629860}

\bibitem[{{Flock} {et~al.}(2015){Flock}, {Ruge}, {Dzyurkevich}, {Henning}, {Klahr}, \& {Wolf}}]{Flock-2015}
{Flock}, M., {Ruge}, J.~P., {Dzyurkevich}, N., {et~al.} 2015, \aap, 574, A68, \dodoi{10.1051/0004-6361/201424693}

\bibitem[{{Harter} {et~al.}(2020){Harter}, {Ricci}, {Zhang}, \& {Zhu}}]{Harter2020}
{Harter}, S.~K., {Ricci}, L., {Zhang}, S., \& {Zhu}, Z. 2020, \apj, 905, 24, \dodoi{10.3847/1538-4357/abcafc}

\bibitem[{{Huang} {et~al.}(2018){Huang}, {Andrews}, {Dullemond}, {Isella}, {P{\'e}rez}, {Guzm{\'a}n}, {{\"O}berg}, {Zhu}, {Zhang}, {Bai}, {Benisty}, {Birnstiel}, {Carpenter}, {Hughes}, {Ricci}, {Weaver}, \& {Wilner}}]{2018HuangDSHARP2}
{Huang}, J., {Andrews}, S.~M., {Dullemond}, C.~P., {et~al.} 2018, \apjl, 869, L42, \dodoi{10.3847/2041-8213/aaf740}

\bibitem[{{Ilee} {et~al.}(2020){Ilee}, {Hall}, {Walsh}, {Jim{\'e}nez-Serra}, {Pinte}, {Terry}, {Bourke}, \& {Hoare}}]{Ilee_2020}
{Ilee}, J.~D., {Hall}, C., {Walsh}, C., {et~al.} 2020, \mnras, 498, 5116, \dodoi{10.1093/mnras/staa2699}

\bibitem[{{Johansen} {et~al.}(2007){Johansen}, {Oishi}, {Mac Low}, {Klahr}, {Henning}, \& {Youdin}}]{Johansen-2007}
{Johansen}, A., {Oishi}, J.~S., {Mac Low}, M.-M., {et~al.} 2007, \nat, 448, 1022, \dodoi{10.1038/nature06086}

\bibitem[{{Kanagawa} {et~al.}(2016){Kanagawa}, {Muto}, {Tanaka}, {Tanigawa}, {Takeuchi}, {Tsukagoshi}, \& {Momose}}]{Kanagawa-2016}
{Kanagawa}, K.~D., {Muto}, T., {Tanaka}, H., {et~al.} 2016, \pasj, 68, 43, \dodoi{10.1093/pasj/psw037}

\bibitem[{{Keppler} {et~al.}(2018){Keppler}, {Benisty}, {M{\"u}ller}, {Henning}, {van Boekel}, {Cantalloube}, {Ginski}, {van Holstein}, {Maire}, {Pohl}, {Samland}, {Avenhaus}, {Baudino}, {Boccaletti}, {de Boer}, {Bonnefoy}, {Chauvin}, {Desidera}, {Langlois}, {Lazzoni}, {Marleau}, {Mordasini}, {Pawellek}, {Stolker}, {Vigan}, {Zurlo}, {Birnstiel}, {Brandner}, {Feldt}, {Flock}, {Girard}, {Gratton}, {Hagelberg}, {Isella}, {Janson}, {Juhasz}, {Kemmer}, {Kral}, {Lagrange}, {Launhardt}, {Matter}, {M{\'e}nard}, {Milli}, {Molli{\`e}re}, {Olofsson}, {P{\'e}rez}, {Pinilla}, {Pinte}, {Quanz}, {Schmidt}, {Udry}, {Wahhaj}, {Williams}, {Buenzli}, {Cudel}, {Dominik}, {Galicher}, {Kasper}, {Lannier}, {Mesa}, {Mouillet}, {Peretti}, {Perrot}, {Salter}, {Sissa}, {Wildi}, {Abe}, {Antichi}, {Augereau}, {Baruffolo}, {Baudoz}, {Bazzon}, {Beuzit}, {Blanchard}, {Brems}, {Buey}, {De Caprio}, {Carbillet}, {Carle}, {Cascone}, {Cheetham}, {Claudi}, {Costille}, {Delboulb{\'e}}, {Dohlen}, {Fantinel}, {Feautrier}, {Fusco}, {Giro}, {Gluck},
  {Gry}, {Hubin}, {Hugot}, {Jaquet}, {Le Mignant}, {Llored}, {Madec}, {Magnard}, {Martinez}, {Maurel}, {Meyer}, {M{\"o}ller-Nilsson}, {Moulin}, {Mugnier}, {Orign{\'e}}, {Pavlov}, {Perret}, {Petit}, {Pragt}, {Puget}, {Rabou}, {Ramos}, {Rigal}, {Rochat}, {Roelfsema}, {Rousset}, {Roux}, {Salasnich}, {Sauvage}, {Sevin}, {Soenke}, {Stadler}, {Suarez}, {Turatto}, \& {Weber}}]{Keppler-2018}
{Keppler}, M., {Benisty}, M., {M{\"u}ller}, A., {et~al.} 2018, \aap, 617, A44, \dodoi{10.1051/0004-6361/201832957}

\bibitem[{{Kley} \& {Nelson}(2012)}]{Kley-Nelson-2012-araa}
{Kley}, W., \& {Nelson}, R.~P. 2012, \araa, 50, 211, \dodoi{10.1146/annurev-astro-081811-125523}

\bibitem[{{Li} {et~al.}(2001){Li}, {Colgate}, {Wendroff}, \& {Liska}}]{rwi-2001}
{Li}, H., {Colgate}, S.~A., {Wendroff}, B., \& {Liska}, R. 2001, \apj, 551, 874, \dodoi{10.1086/320241}

\bibitem[{{Liu}(2019)}]{Baobab_2019}
{Liu}, H.~B. 2019, \apjl, 877, L22, \dodoi{10.3847/2041-8213/ab1f8e}

\bibitem[{{Liu} {et~al.}(2024){Liu}, {Muto}, {Konishi}, {Chung}, {Hashimoto}, {Doi}, {Dong}, {Kudo}, {Hasegawa}, {Terada}, \& {Kataoka}}]{Baobab_2024}
{Liu}, H.~B., {Muto}, T., {Konishi}, M., {et~al.} 2024, arXiv e-prints, arXiv:2402.02900, \dodoi{10.48550/arXiv.2402.02900}

\bibitem[{{Loinard} {et~al.}(2007){Loinard}, {Torres}, {Mioduszewski}, {Rodr{\'\i}guez}, {Gonz{\'a}lez-L{\'o}pezlira}, {Lachaume}, {V{\'a}zquez}, \& {Gonz{\'a}lez}}]{Taurus-2007}
{Loinard}, L., {Torres}, R.~M., {Mioduszewski}, A.~J., {et~al.} 2007, \apj, 671, 546, \dodoi{10.1086/522493}

\bibitem[{{Lombardi} {et~al.}(2008){Lombardi}, {Lada}, \& {Alves}}]{Lupus-2008}
{Lombardi}, M., {Lada}, C.~J., \& {Alves}, J. 2008, \aap, 480, 785, \dodoi{10.1051/0004-6361:20079110}

\bibitem[{{Long} {et~al.}(2019){Long}, {Herczeg}, {Harsono}, {Pinilla}, {Tazzari}, {Manara}, {Pascucci}, {Cabrit}, {Nisini}, {Johnstone}, {Edwards}, {Salyk}, {Menard}, {Lodato}, {Boehler}, {Mace}, {Liu}, {Mulders}, {Hendler}, {Ragusa}, {Fischer}, {Banzatti}, {Rigliaco}, {van de Plas}, {Dipierro}, {Gully-Santiago}, \& {Lopez-Valdivia}}]{Long-2019}
{Long}, F., {Herczeg}, G.~J., {Harsono}, D., {et~al.} 2019, \apj, 882, 49, \dodoi{10.3847/1538-4357/ab2d2d}

\bibitem[{{Long} {et~al.}(2020){Long}, {Pinilla}, {Herczeg}, {Andrews}, {Harsono}, {Johnstone}, {Ragusa}, {Pascucci}, {Wilner}, {Hendler}, {Jennings}, {Liu}, {Lodato}, {Menard}, {van de Plas}, \& {Dipierro}}]{Long-2020}
{Long}, F., {Pinilla}, P., {Herczeg}, G.~J., {et~al.} 2020, \apj, 898, 36, \dodoi{10.3847/1538-4357/ab9a54}

\bibitem[{{Long} {et~al.}(2022){Long}, {Andrews}, {Zhang}, {Qi}, {Benisty}, {Facchini}, {Isella}, {Wilner}, {Bae}, {Huang}, {Loomis}, {{\"O}berg}, \& {Zhu}}]{Long-LkCa15-2022}
{Long}, F., {Andrews}, S.~M., {Zhang}, S., {et~al.} 2022, \apjl, 937, L1, \dodoi{10.3847/2041-8213/ac8b10}

\bibitem[{{Long} {et~al.}(2023){Long}, {Ren}, {Wallack}, {Harsono}, {Herczeg}, {Pinilla}, {Mawet}, {Liu}, {Andrews}, {Bai}, {Cabrit}, {Cieza}, {Johnstone}, {Leisenring}, {Lodato}, {Liu}, {Manara}, {Mulders}, {Ragusa}, {Sallum}, {Shi}, {Tazzari}, {Uyama}, {Wagner}, {Wilner}, \& {Xuan}}]{Long-2023}
{Long}, F., {Ren}, B.~B., {Wallack}, N.~L., {et~al.} 2023, \apj, 949, 27, \dodoi{10.3847/1538-4357/acc843}

\bibitem[{{Lovelace} {et~al.}(1999){Lovelace}, {Li}, {Colgate}, \& {Nelson}}]{RWI-1999}
{Lovelace}, R.~V.~E., {Li}, H., {Colgate}, S.~A., \& {Nelson}, A.~F. 1999, \apj, 513, 805, \dodoi{10.1086/306900}

\bibitem[{{Lynden-Bell} \& {Pringle}(1974)}]{Lynden-Bell_1974}
{Lynden-Bell}, D., \& {Pringle}, J.~E. 1974, \mnras, 168, 603, \dodoi{10.1093/mnras/168.3.603}

\bibitem[{{Mac{\'\i}as} {et~al.}(2021){Mac{\'\i}as}, {Guerra-Alvarado}, {Carrasco-Gonz{\'a}lez}, {Ribas}, {Espaillat}, {Huang}, \& {Andrews}}]{Macias-2021}
{Mac{\'\i}as}, E., {Guerra-Alvarado}, O., {Carrasco-Gonz{\'a}lez}, C., {et~al.} 2021, \aap, 648, A33, \dodoi{10.1051/0004-6361/202039812}

\bibitem[{{Mac{\'\i}as} {et~al.}(2019){Mac{\'\i}as}, {Espaillat}, {Osorio}, {Anglada}, {Torrelles}, {Carrasco-Gonz{\'a}lez}, {Flock}, {Linz}, {Bertrang}, {Henning}, {G{\'o}mez}, {Calvet}, \& {Dent}}]{Macias-2019}
{Mac{\'\i}as}, E., {Espaillat}, C.~C., {Osorio}, M., {et~al.} 2019, \apj, 881, 159, \dodoi{10.3847/1538-4357/ab31a2}

\bibitem[{{Masset}(2000)}]{FARGO}
{Masset}, F. 2000, \aaps, 141, 165, \dodoi{10.1051/aas:2000116}

\bibitem[{{Murphy} {et~al.}(2018){Murphy}, {Bolatto}, {Chatterjee}, {Casey}, {Chomiuk}, {Dale}, {de Pater}, {Dickinson}, {Francesco}, {Hallinan}, {Isella}, {Kohno}, {Kulkarni}, {Lang}, {Lazio}, {Leroy}, {Loinard}, {Maccarone}, {Matthews}, {Osten}, {Reid}, {Riechers}, {Sakai}, {Walter}, \& {Wilner}}]{Murphy-2018-ngVLA}
{Murphy}, E.~J., {Bolatto}, A., {Chatterjee}, S., {et~al.} 2018, in Astronomical Society of the Pacific Conference Series, Vol. 517, Science with a Next Generation Very Large Array, ed. E.~{Murphy}, 3, \dodoi{10.48550/arXiv.1810.07524}

\bibitem[{{Nayakshin} {et~al.}(2020){Nayakshin}, {Tsukagoshi}, {Hall}, {Vazan}, {Helled}, {Humphries}, {Meru}, {Neunteufel}, \& {Panic}}]{Nayakshin-2020}
{Nayakshin}, S., {Tsukagoshi}, T., {Hall}, C., {et~al.} 2020, \mnras, 495, 285, \dodoi{10.1093/mnras/staa1132}

\bibitem[{{Okuzumi} {et~al.}(2016){Okuzumi}, {Momose}, {Sirono}, {Kobayashi}, \& {Tanaka}}]{Okuzumi-2016}
{Okuzumi}, S., {Momose}, M., {Sirono}, S.-i., {Kobayashi}, H., \& {Tanaka}, H. 2016, \apj, 821, 82, \dodoi{10.3847/0004-637X/821/2/82}

\bibitem[{{Ortiz-Le{\'o}n} {et~al.}(2017){Ortiz-Le{\'o}n}, {Loinard}, {Kounkel}, {Dzib}, {Mioduszewski}, {Rodr{\'\i}guez}, {Torres}, {Gonz{\'a}lez-L{\'o}pezlira}, {Pech}, {Rivera}, {Hartmann}, {Boden}, {Evans}, {Brice{\~n}o}, {Tobin}, {Galli}, \& {Gudehus}}]{Ophiuchus-2017}
{Ortiz-Le{\'o}n}, G.~N., {Loinard}, L., {Kounkel}, M.~A., {et~al.} 2017, \apj, 834, 141, \dodoi{10.3847/1538-4357/834/2/141}

\bibitem[{{Pinte} {et~al.}(2019){Pinte}, {van der Plas}, {M{\'e}nard}, {Price}, {Christiaens}, {Hill}, {Mentiplay}, {Ginski}, {Choquet}, {Boehler}, {Duch{\^e}ne}, {Perez}, \& {Casassus}}]{Pinte-2019}
{Pinte}, C., {van der Plas}, G., {M{\'e}nard}, F., {et~al.} 2019, Nature Astronomy, 3, 1109, \dodoi{10.1038/s41550-019-0852-6}

\bibitem[{{Pinte} {et~al.}(2020){Pinte}, {Price}, {M{\'e}nard}, {Duch{\^e}ne}, {Christiaens}, {Andrews}, {Huang}, {Hill}, {van der Plas}, {Perez}, {Isella}, {Boehler}, {Dent}, {Mentiplay}, \& {Loomis}}]{Pinte-2020}
{Pinte}, C., {Price}, D.~J., {M{\'e}nard}, F., {et~al.} 2020, \apjl, 890, L9, \dodoi{10.3847/2041-8213/ab6dda}

\bibitem[{{Pinte} {et~al.}(2023){Pinte}, {Hammond}, {Price}, {Christiaens}, {Andrews}, {Chauvin}, {P{\'e}rez}, {Jorquera}, {Garg}, {Norfolk}, {Calcino}, \& {Bonnefoy}}]{Pinte-2023}
{Pinte}, C., {Hammond}, I., {Price}, D.~J., {et~al.} 2023, \mnras, 526, L41, \dodoi{10.1093/mnrasl/slad010}

\bibitem[{{Ricci} {et~al.}(2021){Ricci}, {Harter}, {Ercolano}, \& {Weber}}]{Ricci-2021}
{Ricci}, L., {Harter}, S.~K., {Ercolano}, B., \& {Weber}, M. 2021, \apj, 913, 122, \dodoi{10.3847/1538-4357/abf5d8}

\bibitem[{{Ricci} {et~al.}(2018){Ricci}, {Liu}, {Isella}, \& {Li}}]{Ricci_2018}
{Ricci}, L., {Liu}, S.-F., {Isella}, A., \& {Li}, H. 2018, \apj, 853, 110, \dodoi{10.3847/1538-4357/aaa546}

\bibitem[{{Ricci} {et~al.}(2010){Ricci}, {Testi}, {Natta}, {Neri}, {Cabrit}, \& {Herczeg}}]{Ricci_2010}
{Ricci}, L., {Testi}, L., {Natta}, A., {et~al.} 2010, \aap, 512, A15, \dodoi{10.1051/0004-6361/200913403}

\bibitem[{{Riols} \& {Lesur}(2019)}]{Riols-2019}
{Riols}, A., \& {Lesur}, G. 2019, \aap, 625, A108, \dodoi{10.1051/0004-6361/201834813}

\bibitem[{{Riols} {et~al.}(2020){Riols}, {Lesur}, \& {Menard}}]{Riols-2020}
{Riols}, A., {Lesur}, G., \& {Menard}, F. 2020, \aap, 639, A95, \dodoi{10.1051/0004-6361/201937418}

\bibitem[{{Savage} \& {Mathis}(1979)}]{Savage-Mathis-1979}
{Savage}, B.~D., \& {Mathis}, J.~S. 1979, \araa, 17, 73, \dodoi{10.1146/annurev.aa.17.090179.000445}

\bibitem[{{Semenov} {et~al.}(2003){Semenov}, {Henning}, {Helling}, {Ilgner}, \& {Sedlmayr}}]{Semenov-2003}
{Semenov}, D., {Henning}, T., {Helling}, C., {Ilgner}, M., \& {Sedlmayr}, E. 2003, \aap, 410, 611, \dodoi{10.1051/0004-6361:20031279}

\bibitem[{{Shakura} \& {Sunyaev}(1973)}]{SS_1973}
{Shakura}, N.~I., \& {Sunyaev}, R.~A. 1973, \aap, 500, 33

\bibitem[{{Suriano} {et~al.}(2018){Suriano}, {Li}, {Krasnopolsky}, \& {Shang}}]{Suriano-2018}
{Suriano}, S.~S., {Li}, Z.-Y., {Krasnopolsky}, R., \& {Shang}, H. 2018, \mnras, 477, 1239, \dodoi{10.1093/mnras/sty717}

\bibitem[{{Teague} {et~al.}(2018){Teague}, {Bae}, {Bergin}, {Birnstiel}, \& {Foreman-Mackey}}]{Teague-2018}
{Teague}, R., {Bae}, J., {Bergin}, E.~A., {Birnstiel}, T., \& {Foreman-Mackey}, D. 2018, \apjl, 860, L12, \dodoi{10.3847/2041-8213/aac6d7}

\bibitem[{{Tsukagoshi} {et~al.}(2019){Tsukagoshi}, {Muto}, {Nomura}, {Kawabe}, {Kanagawa}, {Okuzumi}, {Ida}, {Walsh}, {Millar}, {Takahashi}, {Hashimoto}, {Uyama}, \& {Tamura}}]{Tsukagoshi-2019}
{Tsukagoshi}, T., {Muto}, T., {Nomura}, H., {et~al.} 2019, \apjl, 878, L8, \dodoi{10.3847/2041-8213/ab224c}

\bibitem[{{Ueda} {et~al.}(2022){Ueda}, {Ricci}, {Flock}, \& {Castro}}]{Ueda-2022}
{Ueda}, T., {Ricci}, L., {Flock}, M., \& {Castro}, Z. 2022, \apj, 928, 110, \dodoi{10.3847/1538-4357/ac56d8}

\bibitem[{{Weidenschilling}(1977)}]{Weidenschilling1977}
{Weidenschilling}, S.~J. 1977, \mnras, 180, 57, \dodoi{10.1093/mnras/180.2.57}

\bibitem[{{Whittet} {et~al.}(1997){Whittet}, {Prusti}, {Franco}, {Gerakines}, {Kilkenny}, {Larson}, \& {Wesselius}}]{Chamaeleon-1997}
{Whittet}, D.~C.~B., {Prusti}, T., {Franco}, G.~A.~P., {et~al.} 1997, \aap, 327, 1194

\bibitem[{{Wu} {et~al.}(2023{\natexlab{a}}){Wu}, {Baruteau}, \& {Nayakshin}}]{Wu2023}
{Wu}, Y., {Baruteau}, C., \& {Nayakshin}, S. 2023{\natexlab{a}}, \mnras, 523, 4869, \dodoi{10.1093/mnras/stad1791}

\bibitem[{{Wu} {et~al.}(2023{\natexlab{b}}){Wu}, {Chen}, {Jiang}, {Dong}, {Mac{\'\i}as}, {Lin}, {Rosotti}, \& {Elbakyan}}]{Wu-Chen-Jiang_2023}
{Wu}, Y., {Chen}, Y.-X., {Jiang}, H., {et~al.} 2023{\natexlab{b}}, \mnras, 523, 2630, \dodoi{10.1093/mnras/stad1553}

\bibitem[{{Wu} {et~al.}(2024{\natexlab{a}}){Wu}, {Chen}, \& {Lin}}]{Wu-Chen-Lin-2024}
{Wu}, Y., {Chen}, Y.-X., \& {Lin}, D. N.~C. 2024{\natexlab{a}}, \mnras, 528, L127, \dodoi{10.1093/mnrasl/slad183}

\bibitem[{{Wu} {et~al.}(2024{\natexlab{b}}){Wu}, {Lin}, {Cui}, {Krapp}, {Lee}, \& {Youdin}}]{Wu_Lin-2024}
{Wu}, Y., {Lin}, M.-K., {Cui}, C., {et~al.} 2024{\natexlab{b}}, \apj, 962, 173, \dodoi{10.3847/1538-4357/ad15fe}

\bibitem[{{Youdin} \& {Goodman}(2005)}]{Youdin-2005}
{Youdin}, A.~N., \& {Goodman}, J. 2005, \apj, 620, 459, \dodoi{10.1086/426895}

\bibitem[{{Zhang} {et~al.}(2015){Zhang}, {Blake}, \& {Bergin}}]{Zhang-2015}
{Zhang}, K., {Blake}, G.~A., \& {Bergin}, E.~A. 2015, \apjl, 806, L7, \dodoi{10.1088/2041-8205/806/1/L7}

\bibitem[{{Zhang} {et~al.}(2018){Zhang}, {Zhu}, {Huang}, {Guzm{\'a}n}, {Andrews}, {Birnstiel}, {Dullemond}, {Carpenter}, {Isella}, {P{\'e}rez}, {Benisty}, {Wilner}, {Baruteau}, {Bai}, \& {Ricci}}]{DSHARP-VII}
{Zhang}, S., {Zhu}, Z., {Huang}, J., {et~al.} 2018, \apjl, 869, L47, \dodoi{10.3847/2041-8213/aaf744}

\end{thebibliography}
\bibliographystyle{aasjournal}

\end{CJK*}
\end{document}